\documentclass[camera,letterpaper,nomarginnotes,nonarrowgutter]{jpaper}

\usepackage{dblfloatfix}
\usepackage{multirow}
\usepackage{booktabs}
\usepackage{array}
\usepackage{cite}
\usepackage{algorithm}
\usepackage[noend]{algpseudocode}
\usepackage[rightComments=true,commentColor=gray]{algpseudocodex}
\usepackage{eucal}
\usepackage{rotating}
\usepackage{xcolor,colortbl}
\definecolor{freakishgreen}{HTML}{0A982B}
\definecolor{urlblue}{HTML}{319dd6}
\usepackage{ifthen} 
\usepackage{xspace} 
\usepackage{enumitem}
\usepackage{fancyhdr}
\usepackage{enumitem}
\usepackage[noend]{algpseudocode}
\usepackage{datetime}
\usepackage{textcomp} 
\usepackage{xparse} 
\usepackage{setspace} 
\usepackage{listings} 
\newcolumntype{L}[1]{>{\raggedright\let\newline\\\arraybackslash\hspace{0pt}}m{#1}}
\newcolumntype{C}[1]{>{\centering\let\newline\\\arraybackslash\hspace{0pt}}m{#1}}
\newcolumntype{R}[1]{>{\raggedleft\let\newline\\\arraybackslash\hspace{0pt}}m{#1}}
\newcolumntype{M}[1]{>{\centering\arraybackslash}m{#1}}

\usepackage{tikz}
\newcommand*\circled[1]{\tikz[baseline=(char.base)]{
            \node[shape=circle,fill,inner sep=1pt] (char) {\textcolor{white}{#1}};}}

\makeatletter
\def\thickhline{%
  \noalign{\ifnum0=`}\fi\hrule \@height \thickarrayrulewidth \futurelet
   \reserved@a\@xthickhline}
\def\@xthickhline{\ifx\reserved@a\thickhline
               \vskip\doublerulesep
               \vskip-\thickarrayrulewidth
             \fi
      \ifnum0=`{\fi}}
\makeatother
\newlength{\thickarrayrulewidth}
\usepackage{pifont}

\newcommand\Tstrut{\rule{0pt}{2ex}}         
\newcommand\Bstrut{\rule[-1ex]{0pt}{0pt}}   
\newcommand{\Tabval}[1]{{\Tstrut #1 \Bstrut}}   

\def\thickhline{\noalign{\hrule height.8pt}}
\newcolumntype{?}{!{\vrule width 0.8pt}}

\newcommand{\paraheading}[1]{\vspace{1.5pt}\noindent \textbf{#1}}
\newcommand{\paraspace}{\vspace{1.5pt}\noindent}

\usepackage{xhfill}

\usepackage{soul}
\setul{0.5ex}{0.3ex}
\setulcolor{blue}

\newif\ifsubmission
\submissiontrue

\newif\ifarxiv
\arxivtrue

\definecolor{colorV1}{rgb}{0.0, 0.0, 1.0}    
\definecolor{colorV2}{rgb}{1.0, 0.0, 1.0}    
\definecolor{colorV3}{rgb}{0.0, 0.5, 0.0}    
\definecolor{colorV4}{rgb}{0.6, 0.4, 0.2}    

\newcounter{CurrentDraftVersion}

\ExplSyntaxOn

\cs_new:Npn \revcol:n #1
  {
    \int_case:nnF {#1}
      {
        {1}{blue}
        {2}{magenta}
        {3}{orange}
        {4}{cyan}
        {5}{violet}
        {6}{teal}
        {7}{purple}
        {8}{brown}
        {9}{olive}
        {10}{red}
      }
      {blue}
  }

\NewDocumentCommand{\edit}{ O{\value{CurrentDraftVersion}} m }
  {
    \int_compare:nNnTF { \value{CurrentDraftVersion} } = {1000}
      { #2 }
      {
        \int_compare:nNnTF { \value{CurrentDraftVersion} } = {100}
          { \textcolor{blue}{#2} }
          {
            \int_compare:nNnTF { #1 } = { \int_eval:n { \value{CurrentDraftVersion} - 1 } }
              { \textcolor{\revcol:n {#1}}{#2} }
              { #2 }
          }
      }
  }

\ExplSyntaxOff

\ifsubmission
    \fancypagestyle{firstpage}
    {
        \fancyhead{}
        \begin{tikzpicture}[remember picture,overlay]
        \node [xshift=170mm,yshift=-10mm]
        at (current page.north west) {\href{https://github.com/ctuning/artifact-evaluation/blob/master/docs/reviewing.md}{\includegraphics[scale=0.29]{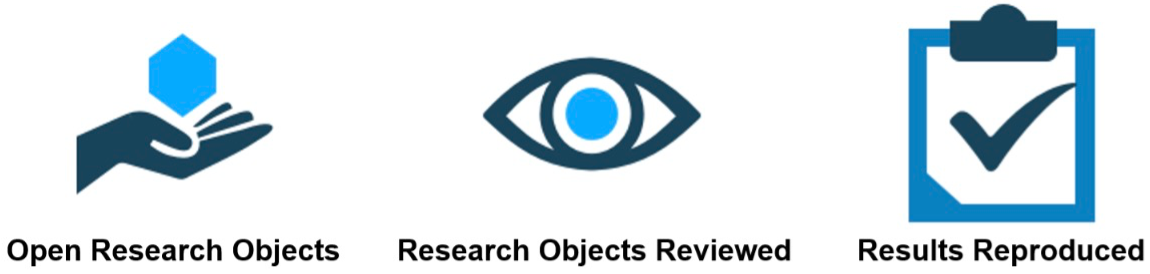}}} ;
        \end{tikzpicture}

        \pagenumbering{arabic}
        \fancyfoot[C]{\large\thepage}
    }
    
\else
    
    \fancypagestyle{firstpage}
    {
        \fancyhead[C]{\textcolor{blue}{\emph{Version \arabic{CurrentDraftVersion}~---~\today, \xxivtime \ UTC}}}
        \fancyfoot[C]{\thepage}
    }
\fi

\usepackage{hyperref}

\hypersetup{
    bookmarks=true,
    breaklinks=true,
    colorlinks=true,
    linkcolor=blue,
    citecolor=blue,
    urlcolor=urlblue,
    pdfpagelabels=false,
}

\usepackage[nameinlink]{cleveref}
\crefformat{section}{\S#2#1#3} 
\crefformat{subsection}{\S#2#1#3}
\crefformat{subsubsection}{\S#2#1#3}

\lstdefinestyle{shell}{
  basicstyle=\ttfamily\small,
  columns=fullflexible,
  keepspaces=true,
  breaklines=true,
  showstringspaces=false,
  upquote=true,
  frame=none,        
  xleftmargin=0pt,   
  aboveskip=6pt,
  belowskip=6pt
}
\def\BibTeX{{\rm B\kern-.05em{\sc i\kern-.025em b}\kern-.08em
    T\kern-.1667em\lower.7ex\hbox{E}\kern-.125emX}}

\title{\LARGE{Athena: Synergizing Data Prefetching and Off-Chip Prediction \\ via Online Reinforcement Learning}}

\author{
    *Rahul Bera$^1$ \hspace{0.5em} *Zhenrong Lang$^1$ \hspace{0.5em} Caroline Hengartner$^1$ \hspace{0.5em} Konstantinos Kanellopoulos$^1$ \vspace{0.2em} \\
    Rakesh Kumar$^2$ \hspace{0.5em} Mohammad Sadrosadati$^1$ \hspace{0.5em} Onur Mutlu$^1$ \vspace{0.5em} \\
    \normalsize{
        $^1$ETH Zürich \hspace{0.5em} $^2$NTNU
    }
}

\begin{document}

\maketitle

\def\thefootnote{}\footnotetext{*Rahul Bera and Zhenrong Lang are co-primary authors.}\def\thefootnote{\arabic{footnote}}

\thispagestyle{firstpage}

\begin{abstract}
Prefetching and off-chip prediction are two techniques proposed to hide long memory access latencies in high-performance processors.
In this work, we demonstrate that: 
(1) prefetching and off-chip prediction often provide complementary performance benefits, 
yet (2) naively combining these two mechanisms often fails to realize their full performance potential, 
and (3) existing prefetcher control policies (both heuristic- and learning-based) leave significant room for performance improvement behind.

Our goal is to design a holistic framework that can autonomously learn to coordinate an off-chip predictor with multiple prefetchers employed at various cache levels, delivering \emph{consistent} performance benefits across a wide range of workloads and system configurations.

To this end, we propose a new technique called Athena, which models the coordination between prefetchers and off-chip predictor (OCP) as a reinforcement learning (RL) problem.
Athena acts as the RL agent that observes multiple system-level features (e.g., prefetcher/OCP accuracy, bandwidth usage) over an epoch of program execution, and uses them as \emph{state} information to select a coordination \emph{action} (i.e., enabling the prefetcher and/or OCP, and adjusting prefetcher aggressiveness). 
At the end of every epoch, Athena receives a numerical \emph{reward} that measures the change in multiple system-level metrics (e.g., number of cycles taken to execute an epoch).
Athena uses this reward to autonomously and continuously learn a policy to coordinate prefetchers with OCP.

Athena makes a key observation that using performance improvement as the sole RL reward, as used in prior work, is unreliable, as it confounds the effects of the agent's actions with inherent variations in workload behavior.
To address this limitation, Athena introduces a composite reward framework that separates 
(1) system-level metrics directly influenced by Athena's actions (e.g., last-level cache misses) from 
(2) metrics primarily driven by workload phase changes (e.g., number of mispredicted branches).
This allows Athena to autonomously learn a coordination policy by isolating the true impact of its actions from inherent variations in workload behavior.

Our extensive evaluation using a diverse set of 100 memory-intensive workloads shows that Athena \emph{consistently} outperforms multiple prior state-of-the-art coordination policies across a wide range of system configurations with various combinations of underlying prefetchers at various cache levels, OCPs, and main memory bandwidths, while incurring only modest storage overhead and design complexity.
The source code of Athena is freely available at \url{https://github.com/CMU-SAFARI/Athena}.
\end{abstract}

\section{Introduction} \label{sec:intro}

\emph{Data prefetching}~\cite{stride, streamer, baer2, stride_vector, jouppi_prefetch, ampm, fdp, footprint, sms, sms_mod, spp, vldp, sandbox, bop, dol, dspatch, mlop, ppf, ipcp, pythia, chen2025gaze, dmp, markov, stems, somogyi_stems, wenisch2010making, domino, isb, misb, triage, wenisch2005temporal, chilimbi2002dynamic, chou2007low, ferdman2007last, hu2003tcp, bekerman1999correlated, cooksey2002stateless, karlsson2000prefetching, ssmt, precompute, software_preexecute, sohi_slice, helper_thread, bfetch, runahead_threads, helper_thread2, zhang2007accelerating, dubois1998assisted, collins2001speculative, solihin2002using, Roth2001SpeculativeDM, Sundaramoorthy2000SlipstreamPI, hashemi2016continuous, Moshovos2001SliceprocessorsAI, Chappell2002DifficultpathBP, zhou2005dual, srinivasan2004continual, Garg2008APE, Parihar2014AcceleratingDL, Purser2000ASO, RothMS99} and \emph{off-chip prediction}~\cite{hermes, lp, krause2022hbpb} are two speculation techniques proposed to hide the long memory access latency in high-performance processors.
A data prefetcher hides memory access latency by predicting full cacheline addresses of memory requests and fetching their corresponding data into on-chip caches before the program demands them.
An off-chip predictor (OCP), on the other hand, predicts which memory requests would go off-chip and starts fetching their data directly from the main memory, thereby hiding the on-chip cache access latency from the critical path of off-chip memory requests.

\paraheading{Key Observations.}
While prior works demonstrate performance benefits of combining OCP with data prefetchers~\cite{hermes, jamet2024tlp}, we make three key observations that highlight significant room for further performance improvement in processors using both prefetchers and OCP.
First, we show that prefetching and off-chip prediction often provide complementary performance benefits, especially in bandwidth-constrained processors, where a prefetcher often \emph{degrades} performance due to its bandwidth overhead~\cite{fdp, clip, ebrahimi_paware, ebrahimi2009coordinated, ebrahimi2009techniques, pythia, micro_mama, lee2008prefetch, memory_bank}, but an OCP \emph{improves} performance due to its accurate predictions.
Second, naively combining prefetching and off-chip prediction often fails to realize their full performance potential together. 
More specifically, in workloads where a prefetcher degrades performance, it often masks the performance gains that an off-chip predictor alone could otherwise deliver.
Third, existing coordination techniques~\cite{mab, ebrahimi2009coordinated, jamet2024tlp} either lack the flexibility in coordinating OCP in the presence of multiple prefetchers employed at various cache levels~\cite{jamet2024tlp}, or leave substantial room for performance improvement behind~\cite{ebrahimi2009coordinated, mab}.

\textbf{Our goal} in this work is to design a holistic framework that can autonomously coordinate off-chip prediction with \emph{multiple} prefetching techniques employed at various levels of the cache hierarchy by taking multiple system-level features into account, thereby delivering \emph{consistent} performance benefits, regardless of the underlying workload and system configuration.

To this end, we propose a new technique called \emph{\textbf{Athena}},\footnote{Named after the Greek goddess of wisdom and strategic warfare~\cite{athena_wiki}.} 
which formulates the dynamic coordination between prefetchers and OCP as a reinforcement learning problem. 
Athena observes a set of system-level features (e.g., accuracy of prefetchers and OCP, main memory bandwidth usage, prefetcher-induced cache pollution, etc.) over an epoch of workload execution. 
At the end of an epoch, Athena uses the feature values as \emph{state} information to take a coordination \emph{action} (i.e., enabling the OCP and/or prefetcher, and adjusting prefetcher aggressiveness).
Athena also receives a numerical \emph{reward} at the end of every epoch, which measures the change in multiple system-level metrics (e.g., number of cycles taken to execute an epoch) to evaluate the impact of its actions on the system. 
Athena uses this reward to autonomously and continuously learn and update a prefetcher-OCP coordination policy, that can adapt to changing workloads and system configurations.

\paraheading{Novelty and Benefits.}
While prior works have explored RL-based prefetcher control (e.g.,~\cite{pythia, jalili2022managing, mab, micro_mama}), Athena is the first work that systematically studies the interactions between state-of-the-art prefetchers and OCP, and provides an effective RL-based coordination policy. 
Athena differs from prior works in three key ways.
First, prior works use the change in instructions committed per cycle (IPC) per epoch as the \emph{only} reward for their RL-based learning mechanisms~\cite{mab, jalili2022managing, micro_mama}.
We observe that evaluating the effectiveness of a prefetcher and/or an OCP solely by the IPC change can be unreliable due to the inherent IPC variations caused by changes in workload behavior.
To mitigate this, Athena introduces a composite reward framework that 
explicitly separates (1) system-level metrics that are directly influenced by Athena's actions (e.g., execution cycles, last-level cache misses) from 
(2) metrics that are primarily driven by the change in workload behavior (e.g., number of mispredicted branch instructions).
This allows Athena to autonomously learn a coordination policy by isolating the true impact of its actions from inherent variations in workload behavior.
We believe that this composite reward structure can be broadly applicable to many other microarchitectural decision-making processes beyond prefetcher-OCP coordination.
Second, Athena uses the expected reward of a selected action not only to coordinate the prefetcher and OCP, but also to dynamically adjust prefetcher aggressiveness. 
Thus, Athena works as a prefetcher-OCP coordinator \emph{and} a prefetcher throttler, at the same time, using the same hardware.
Third, rather than using computationally-intensive deep RL~\cite{jalili2022managing}, or state-agnostic RL methods~\cite{mab, micro_mama}, Athena leverages a lightweight, state-aware SARSA RL algorithm~\cite{sarsa}.
Thus, Athena significantly reduces storage and computational overheads compared to~\cite{jalili2022managing}, while outperforming the lightweight state-agnostic approach~\cite{mab} (see~\cref{sec:eval_sc}).

\paraheading{Results Summary.}
We extensively evaluate Athena using a diverse set of $100$ memory-intensive workloads across a wide range of system configurations by varying 
(1) the cache hierarchy design (i.e., changing the number of prefetchers employed at each cache level), 
(2) prefetcher type, 
(3) off-chip predictor type, and 
(4) the main memory bandwidth. 
We compare Athena against three prior techniques: 
TLP~\cite{jamet2024tlp}, which is proposed to coordinate OCP only with a prefetcher at L1 data cache,
and HPAC~\cite{ebrahimi2009coordinated} and MAB~\cite{mab}, both of which are proposed to coordinate only prefetchers present at various levels of the cache hierarchy.
Our evaluation yields five major findings.
First, in a bandwidth-constrained single-core processor, Athena, coordinating POPET~\cite{hermes} as the OCP and Pythia~\cite{pythia} as the prefetcher at L2 cache (L2C), outperforms the next-best-performing coordination technique, MAB~\cite{mab}, by $5.0\%$ on average.
Second, Athena \emph{consistently} outperforms \emph{all} prior coordination techniques across \emph{all} cache designs. 
Athena outperforms the next-best coordination techniques by $5.2\%$, $6.4\%$, and $7.0\%$ in three different cache designs that employ 
(a) one prefetcher at L1 data cache (L1D), 
(b) two prefetchers at L2 cache (L2C), 
and (c) one prefetcher each at L1D and L2C, respectively.
Third, keeping POPET as the OCP, varying the L1D prefetcher between IPCP~\cite{ipcp} and Berti~\cite{navarro2022berti}, and the L2C prefetcher among SPP+PPF~\cite{spp, ppf}, MLOP~\cite{mlop}, and SMS~\cite{sms}, Athena \emph{consistently outperforms} all prior techniques on average by $3.6\%$-$10.3\%$ across all workloads.
Fourth, keeping Pythia as the L2C prefetcher and varying the OCP, we show that Athena \emph{consistently outperforms} MAB by $4.7\%$ and $8.2\%$ on average across all workloads, while using a hit-miss predictor (HMP~\cite{yoaz1999speculation}) and a tag-tracking-based predictor (TTP~\cite{lp, hermes}) as the underlying OCP.
Fifth, all of Athena's benefits come at a modest storage overhead ($3$~KB per core) and design complexity.

\vspace{2pt}
\paraspace We make the following key contributions in this work:
\vspace{2pt}

\begin{itemize}
\item We demonstrate that, even though off-chip prediction and data prefetching often provide complementary performance benefits, naively combining these two mechanisms often fails to realize their full performance potential.

\item We introduce Athena, a new reinforcement learning (RL)-based technique that exploits various system-level features to autonomously learn to (1) coordinate an off-chip predictor and multiple prefetchers and (2) control prefetcher aggressiveness, at the same time, using the same hardware.

\item Athena introduces a novel, and potentially broadly-applicable, RL reward framework that allows Athena to reliably learn a coordination policy by isolating the true impact of its actions from inherent variations in workload behavior.

\item We show that Athena \emph{consistently} outperforms state-of-the-art prefetcher coordination policies across $100$ diverse workloads, six different prefetchers at different cache levels, three different off-chip predictors, a wide range of memory bandwidth configurations, and varying cache designs.

\item We open-source Athena and all the workload traces used for performance modeling in our GitHub repository: \url{https://github.com/CMU-SAFARI/Athena}.

\end{itemize}
\section{Motivation} 
\label{sec:motivation}

Long memory access latency remains a key performance bottleneck in modern processors~\cite{mutlu2003runahead, mutlu2003runahead2, mutlu2006efficient, patel2024rethinking, lee2013tiered, lee2015adaptive}. 
Researchers have proposed numerous techniques to hide memory latency: data prefetching~\cite{stride, streamer, baer2, stride_vector, jouppi_prefetch, ampm, fdp, footprint, sms, sms_mod, spp, vldp, sandbox, bop, dol, dspatch, mlop, ppf, ipcp, pythia, chen2025gaze, dmp, markov, stems, somogyi_stems, wenisch2010making, domino, isb, misb, triage, wenisch2005temporal, chilimbi2002dynamic, chou2007low, ferdman2007last, hu2003tcp, bekerman1999correlated, cooksey2002stateless, karlsson2000prefetching, ssmt, precompute, software_preexecute, sohi_slice, helper_thread, bfetch, runahead_threads, helper_thread2, zhang2007accelerating, dubois1998assisted, collins2001speculative, solihin2002using, Roth2001SpeculativeDM, Sundaramoorthy2000SlipstreamPI, hashemi2016continuous, Moshovos2001SliceprocessorsAI, Chappell2002DifficultpathBP, zhou2005dual, srinivasan2004continual, Garg2008APE, Parihar2014AcceleratingDL, Purser2000ASO, RothMS99} and off-chip prediction~\cite{hermes, lp, krause2022hbpb} are two such key techniques.

\emph{Data prefetching} is a well-studied speculation technique that predicts addresses of memory requests and fetches their corresponding data into on-chip caches before the processor demands them. 
When accurate, prefetching improves performance by hiding the memory access latency. 
However, incorrect speculation can lead to significant memory bandwidth overhead and cache pollution~\cite{fdp, ebrahimi2009techniques}.
Prior works have shown that prefetchers often lose their performance benefits (and even severely degrade performance) in processors with limited memory bandwidth or cache capacity~\cite{fdp, clip, ebrahimi_paware, ebrahimi2009coordinated, ebrahimi2009techniques, pythia, memory_bank, lee2008prefetch, micro_mama}.

\emph{Off-chip prediction} is a more recently proposed speculation technique that predicts which memory requests would go off-chip and fetches their data directly from main memory~\cite{hermes, lp, krause2022hbpb}. 
Unlike a prefetcher that predicts \emph{full cacheline addresses} of future memory requests, an off-chip predictor (OCP) makes a \emph{binary} prediction on a memory request with \emph{known} cacheline address: speculating whether or not it will access the off-chip main memory.
This key difference allows OCP to often produce more accurate predictions than a prefetcher (see~\Cref{subsubsec:complementary}). 
However, OCP can only hide on-chip cache access latency from the critical path of an off-chip memory request, offering lower timeliness than a prefetcher.

\subsection{Key Observations}

We make three key observations, highlighting room for performance improvement in processors that employ both OCP and prefetchers: 
(1) prefetcher and OCP often provide complementary performance benefits, especially in a bandwidth-constrained processor configuration, 
yet (2) naively combining these two techniques often fails to realize their full performance potential, 
and (3) existing microarchitectural policies are either not capable of coordinating OCP with multiple prefetchers, or leave substantial room for performance gain.

\subsubsection{\textbf{Off-chip prediction and prefetching provide complementary performance benefits}} \label{subsubsec:complementary}
\begin{figure}[!ht]
    \centering
    \includegraphics[width=\columnwidth]{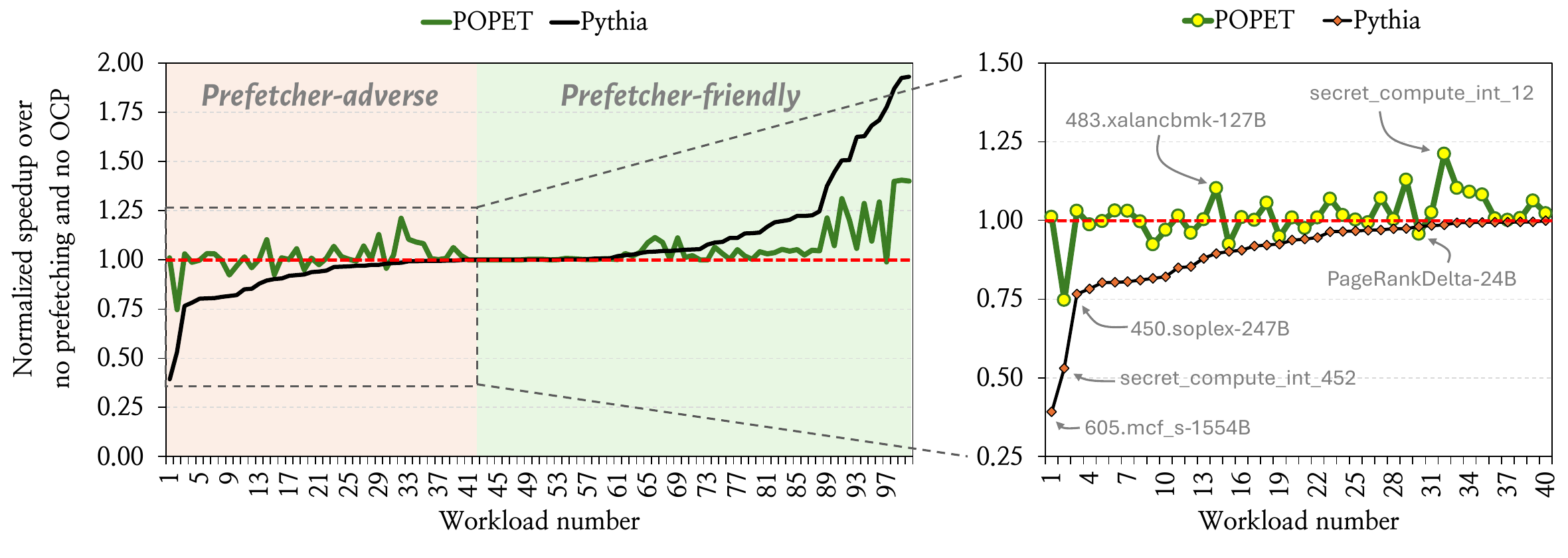}
    \caption{Performance line graph of a state-of-the-art off-chip predictor (OCP), POPET~\cite{hermes}, and a state-of-the-art prefetcher, Pythia~\cite{pythia}, across 100 workloads.}
\label{fig:motiv_perf_line}
\end{figure}

\Cref{fig:motiv_perf_line} shows the performance line graph of a state-of-the-art OCP, POPET~\cite{hermes}, against a state-of-the-art data prefetcher, Pythia~\cite{pythia}, deployed at the L2 cache (L2C) in a memory bandwidth-constrained single-core processor\footnote{We model the memory bandwidth-constrained processor with $3.2$ GB/s of main memory bandwidth (see~{\Cref{sec:methodology}}). 
This configuration closely matches the per-core main memory bandwidth of many commercial datacenter-class processors, e.g., AMD EPYC 9754S~\cite{epyc_9754}, AmpereOne A192~{\cite{ampere_one}}, Amazon Graviton 3~{\cite{graviton3}}, and the Arm Neoverse V2 platform~\cite{bruce2023arm, neoverse2}. Nonetheless, our technique's benefits hold true for a wide range of memory bandwidth configurations, as shown in \Cref{sec:eval_sen_bw}.} across $100$ workloads. 
The graph is sorted in increasing order of Pythia's speedup over the baseline system without a prefetcher or an OCP.

We make three key observations.
First, even though Pythia improves performance for the majority of workloads (highlighted in green), it also \emph{degrades} performance for a significant number of workloads ($40$ out of $100$; highlighted in red) even with its built-in bandwidth-aware throttling mechanism.
For ease of discussion, we call the workloads with performance improvement (or degradation) \emph{prefetcher-friendly} (\emph{prefetcher-adverse}). 
Second, in many prefetcher-adverse workloads, POPET \emph{improves} performance over the baseline.
Pythia \emph{degrades} performance by $11.6\%$ on average across all prefetcher-adverse workloads, whereas POPET \emph{improves} performance by $1.4\%$.
This contrast arises because, in prefetcher-adverse workloads, it is often easier to predict whether a memory request would go off-chip than to predict the full cacheline address of a future memory request.
For instance, in \texttt{483.xalancbmk-127B}, a workload known for its irregular memory access pattern, POPET predicts off-chip requests with $84.1\%$ accuracy, while Pythia generates prefetch requests with only $28.7\%$ accuracy.
As a result, POPET \emph{improves} performance by $10.3\%$, whereas Pythia \emph{degrades} performance by $10.5\%$.
Third, in prefetcher-friendly workloads, however, Pythia provides significantly higher performance benefits ($16.0\%$ on average) than POPET ($5.9\%$ on average). 
This is because, in these workloads, Pythia brings data to the cache well ahead of demand, hiding more memory access latency than only hiding the on-chip cache access latency by POPET.\footnote{Even though we demonstrate this observation using POPET and Pythia as the OCP and prefetcher, respectively, we observe this dichotomy across various prefetcher and OCP implementations. 
In~\Cref{sec:evaluation}, we extend this observation to six prefetcher types~\cite{ipcp, navarro2022berti, pythia, spp, ppf, sms, mlop} and three OCP types~\cite{lp, hermes, yoaz1999speculation}.}

We conclude that OCP and prefetching often provide complementary performance benefits due to their fundamentally different forms of speculation.

\subsubsection{\textbf{Naively combining OCP with prefetching often fails to realize their full performance potential}} \label{subsubsec:naive_combo}
\begin{figure}[!ht]
    \centering
    \includegraphics[width=\columnwidth]{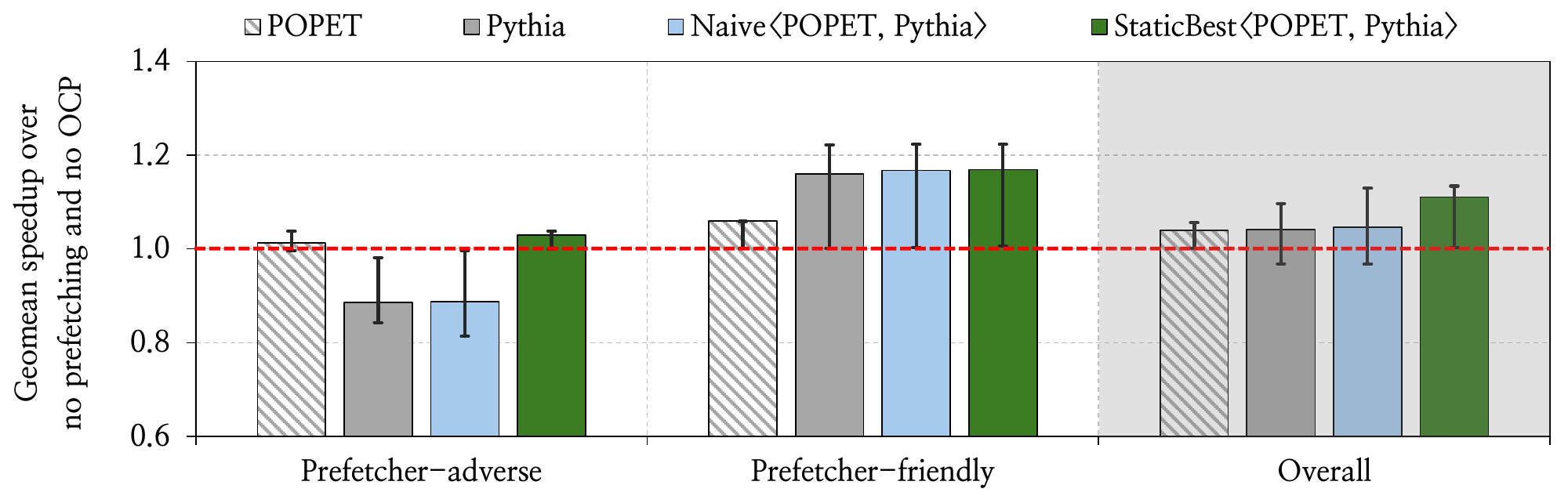}
    \caption{Geomean speedup of POPET, Pythia, Naive, and StaticBest combinations across all workloads.}
\label{fig:motiv2}
\end{figure}

Although OCP and prefetching offer different tradeoffs for different workload categories, naively combining the two mechanisms often fails to realize their full performance potential together.
\Cref{fig:motiv2} compares the performance of POPET and Pythia individually against two combinations of them: 
(1) \emph{Naive<POPET, Pythia>}, that simultaneously enables both POPET and Pythia without any coordination, 
and (2) \emph{StaticBest<POPET, Pythia>}, that retrospectively (i.e., using end-to-end workload execution results offline) selects the best-performing option for each workload among four possibilities: POPET-only, Pythia-only, both enabled, and both disabled.\footnote{While StaticBest estimates the performance headroom of an intelligent coordination policy, it is possible to further improve upon StaticBest by \emph{dynamically} identifying the best combination for \emph{each workload phase}, rather than the entire end-to-end workload.
In~\cref{sec:eval_case_study} we show that, by adapting to each workload phase, our proposed technique can outperform StaticBest.} 
The error bar indicates the range between the first and third quartiles.

We make two key observations from~\Cref{fig:motiv2}. 
First, even though Naive provides $4.7\%$ performance improvement over the baseline across \emph{all} workloads, Naive \emph{degrades} performance by $11.2\%$ in prefetcher-adverse workloads, effectively masking the performance improvement that POPET alone could have delivered otherwise (i.e., $1.2\%$ on average).
This shows that, even though off-chip prediction provides complementary performance benefits to prefetching, especially in prefetcher-adverse workloads, naively combining both techniques does not realize their full performance potential together.
Second, StaticBest combination provides \emph{consistent} performance benefits in \emph{both} prefetcher-adverse and prefetcher-friendly workloads and significantly outperforms the Naive combination (by $6.5\%$ on average) across all workloads.
These observations demonstrate the need to design an intelligent coordination mechanism between OCPs and prefetchers.

\subsubsection{\textbf{Existing coordination policies are either inflexible or leave a large performance potential behind}} \label{subsubsec:prior}
\begin{figure}[!ht]
    \centering
    \includegraphics[width=\columnwidth]{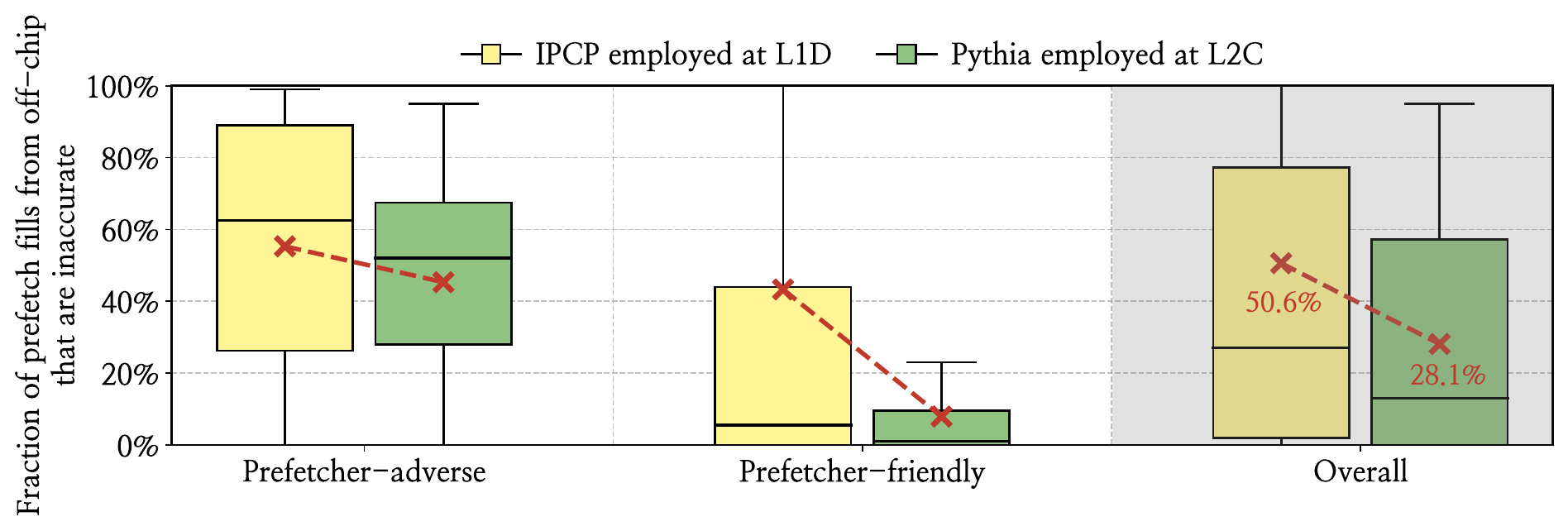}
    \caption{Fraction of prefetch fills from off-chip main memory that are inaccurate.}
\label{fig:tlp_motiv}
\end{figure}

While researchers have proposed numerous techniques to control multiple prefetchers (e.g., ~\cite{ibm_power7, fdp, ebrahimi2009coordinated, ebrahimi2009techniques, eris2022puppeteer, mab, jalili2022managing, yang2024rl, yang2025reinforcement, US20080243268A1, US8583894B2, US9292447B2, US10073785B2, US8156287B2, US9645935B2, US8924651B2, US8892822B2, US9904624B1, US10331567B1}), TLP~\cite{jamet2024tlp} is the \emph{only} prior technique that aims to control a prefetcher in the presence of an OCP.
TLP uses off-chip prediction as a hint to filter out prefetch requests to the L1 data cache (L1D), based on the empirical observation that prefetches filled from off-chip main memory into L1D are often inaccurate (i.e., the cacheline is not subsequently demanded during its cache residency)~\cite{jamet2024tlp}.
While TLP's observation is often effective for an L1D prefetcher, we observe that it may not hold true for prefetchers employed at higher (i.e., further away from the core) cache levels.
\Cref{fig:tlp_motiv} shows the fraction of prefetch fills from the off-chip main memory that are inaccurate as a box-and-whisker plot.\footnote{Each box is lower- (upper-) bounded by the first (third) quartile. 
The box size represents the inter-quartile range (IQR). The whiskers extend to $1.5\times$ IQR range on each side. 
The cross-marked value in the box shows the mean.}
We show the fraction for two state-of-the-art prefetchers, IPCP~\cite{ipcp} and Pythia, individually employed at two different cache levels. 
IPCP fills prefetch requests to L1D, whereas Pythia fills to L2C.
The key observation is that, while $50.6\%$ prefetch fills to L1D caused by IPCP are inaccurate, only $28.1\%$ of the prefetch fills to L2C caused by Pythia are inaccurate. 
In other words, an off-chip prefetch fill to L2C is \emph{nearly half as likely} to be inaccurate as an off-chip prefetch fill to L1D, while employing state-of-the-art prefetchers.
This fundamental limitation in TLP's key observation significantly limits its ability to coordinate OCPs with prefetchers that are placed beyond L1D (as we demonstrate in~\Cref{subsubsec:L1D} and~\Cref{subsubsec:L1DL2C}).

\begin{figure}[!ht]
    \centering
    \includegraphics[width=\columnwidth]{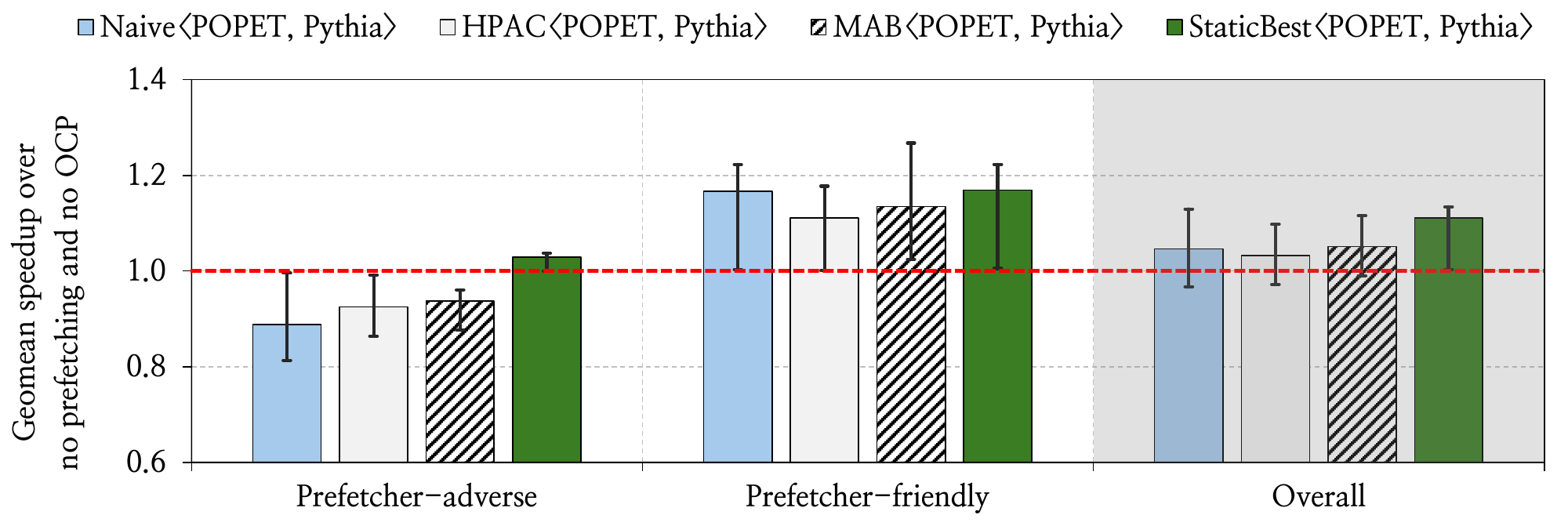}
    \caption{Geomean speedup of Naive, HPAC, MAB, and StaticBest combinations across all workloads.}
\label{fig:motiv3}
\end{figure}

Besides TLP, other prior techniques focus solely on prefetcher control, without considering OCP. 
We \emph{extend} two best-performing prior techniques, heuristic-based HPAC~\cite{ebrahimi2009coordinated} and learning-based MAB~\cite{mab}, to coordinate an OCP and a prefetcher.
\Cref{fig:motiv3} compares the performance of these techniques, coordinating POPET as the OCP and Pythia as the L2C prefetcher against Naive and StaticBest combinations across all workloads.\footnote{\Cref{fig:motiv3} excludes TLP as its key observation does not reliably extend to coordinating an L2C prefetcher (Pythia) with an OCP, as shown earlier.}
We make two key observations.
First, in prefetcher-adverse workloads, both HPAC and MAB considerably mitigate the performance degradation of the Naive combination. However, neither policy matches the performance of the baseline (without prefetching or OCP), let alone harnesses the potential performance gains of StaticBest.
Second, in prefetcher-friendly workloads, these coordination techniques fall short of the Naive combination.
The heuristic-based HPAC falls short due to its reliance on the statically tuned thresholds that are optimized for average-case behavior across workloads. 
These fixed thresholds cannot adapt to per-workload or phase-specific characteristics, causing HPAC to make conservative coordination decisions even when prefetching is beneficial.
While MAB avoids such static thresholds, it still falls short as it makes decisions agnostic to any system-level features (e.g., prefetcher/OCP accuracy, prefetch-induced cache pollution).

We conclude that while there is a rich literature on prefetcher coordination techniques, the only OCP-aware prefetcher control mechanism (i.e., TLP) lacks flexibility (i.e., the ability to coordinate OCP with multiple prefetchers employed at various levels of the cache hierarchy), while other techniques (e.g., HPAC, MAB) leave significant performance potential behind.

\subsection{Our Goal and Proposal} \label{subsec:goal}

\textbf{Our goal} in this work is to design a holistic framework that can autonomously coordinate off-chip prediction with \emph{multiple} prefetching techniques employed at various levels of the cache hierarchy by taking multiple system-level features into account, thereby delivering \emph{consistent} performance benefits, regardless of the underlying prefetcher-OCP combination, workload, and system configuration.

To this end, we propose \emph{\textbf{Athena}}, which formulates the dynamic coordination between prefetchers and OCP as a reinforcement learning problem.
\section{Background}
\label{sec:background}

\subsection{Reinforcement Learning}
\label{subsec:bg_rl}
Reinforcement learning (RL)~\cite{rl_bible, rlmc} is a machine learning paradigm where an agent autonomously learns to make decisions by interacting with its environment. 
A typical RL system consists of two key components: the agent and the environment.
At each timestep $t$, the agent observes the current \emph{\textbf{state}} of the environment $S_t$, selects an \emph{\textbf{action}} $A_t$, and receives a numerical \emph{\textbf{reward}} $R_{t+1}$ based on the outcome of the action. 
The agent's goal is to find the optimal \emph{\textbf{policy}} that maximizes the cumulative reward collected over time. 
The expected cumulative reward for taking an action $A$ in a given state $S$ is defined as the \emph{Q-value} of the state-action pair, denoted as $Q(S, A)$.
At every timestep, the agent iteratively optimizes its policy in two steps: 
(1) update the Q-value using the reward collected from the environment, 
and (2) optimize the policy.

\paraheading{Q-value Update.}
If at a given timestep $t$, the agent observes a state $S_t$, takes an action $A_t$, while the environment transitions to a new state $S_{t+1}$ and emits a reward $R_{t+1}$, and the agent takes action $A_{t+1}$ in the new state, the Q-value of the old state-action pair $Q(S_t, A_t)$ is iteratively optimized using the SARSA algorithm~\cite{sarsa, rl_bible} as:
\begin{equation}\label{eq:sarsa}
\begin{aligned}
Q\left(S_t, A_t\right) & \gets Q\left(S_t, A_t\right)\\
&+ \alpha\left[R_{t+1}+\gamma Q\left(S_{t+1}, A_{t+1}\right) - Q\left(S_t, A_t\right)\right]
\end{aligned}
\end{equation} 
Here, $\alpha$, the \emph{learning rate}, determines the extent to which new information overrides prior knowledge. $\gamma$ is the \emph{discount factor}, which trades off between immediate and future rewards.

\paraheading{Policy Optimization.}
To identify a policy that maximizes the cumulative reward, a greedy agent always selects the action with the highest Q-value in a given state. 
However, such exploitation risks under-exploring the state-action space. 
Thus, to strike a balance between exploration and exploitation, an $\epsilon$-greedy agent \emph{stochastically} takes a random action with a low probability of $\epsilon$ (called \emph{exploration rate}); otherwise, it selects the action that provides the highest Q-value~\cite{rl_bible}.

\subsection{Why is RL a Good Fit for Coordinating Prefetching and Off-Chip Prediction?}
RL is well-suited for coordinating data prefetchers and OCP due to the following three key advantages.

\paraheading{Less reliance on static heuristics and thresholds.}
Heuristic-based prefetcher coordination policies typically rely on statically defined thresholds~\cite{fdp, ebrahimi_fst, ebrahimi2009coordinated, ebrahimi2009techniques, ebrahimi_paware}. 
These thresholds are manually tuned and inherently inflexible, often resulting in suboptimal performance when workloads or system conditions change~\cite{rlmc, morse}.
Formulating prefetcher-OCP coordination as an RL problem allows a hardware architect to focus on \textit{what} performance targets the coordinator should achieve and \textit{which} system-level features might be useful, rather than spending time on manually devising fixed algorithms and/or thresholds that describe \textit{precisely how} the coordinator should achieve that target. 
This not only significantly reduces the human effort needed for a coordinator design, but also yields higher-performing coordination (as shown in~\Cref{sec:evaluation}).

\paraheading{Online feedback-driven learning.}
RL provides two key benefits over supervised learning methods (e.g., SVM~\cite{svm1,svm2}, decision tree~\cite{decision_tree1}) in formulating the prefetcher-OCP coordination problem. 
First, prefetcher-OCP coordination lacks a well-defined ground-truth label (e.g., which mechanism is beneficial to enable) for each system state. 
The optimal coordination decision depends on delayed, system-level performance outcomes that manifest only after executing an action under dynamic resource contention and changing workload behavior. 
As a result, generating labeled training data would require exhaustive offline exploration across workloads, phases, and system configurations, and such labels may not generalize as the underlying hardware, prefetchers, or OCPs change. 
Second, supervised learning models are inherently static once trained and cannot naturally adapt online to changing workload behavior without repeated retraining. 
In contrast, RL directly optimizes long-term performance using online reward feedback and continuously updates its policy, making it better suited to model prefetcher-OCP coordination.

\paraheading{Q-value-driven prefetcher aggressiveness control.}
The Q-values learned by the RL agent provide a natural ranking over available actions, reflecting their expected utility. 
For actions involving prefetching (either standalone or combined with OCP), the corresponding Q-values \emph{implicitly} encode the agent's confidence in the prefetcher's effectiveness relative to alternative actions.
An RL agent can leverage these Q-values to also dynamically control prefetcher aggressiveness. 
Higher Q-values correspond to stronger confidence, prompting more aggressive prefetching, whereas lower Q-values imply uncertain benefits, resulting in more conservative prefetching.
Importantly, this Q-value-driven prefetcher aggressiveness control incurs no additional hardware overhead, since the learned Q-values jointly govern both prefetcher/OCP selection and prefetcher aggressiveness.
\section{Athena: Overview}

\begin{figure}[!ht]
    \centering
    \includegraphics[width=\columnwidth]{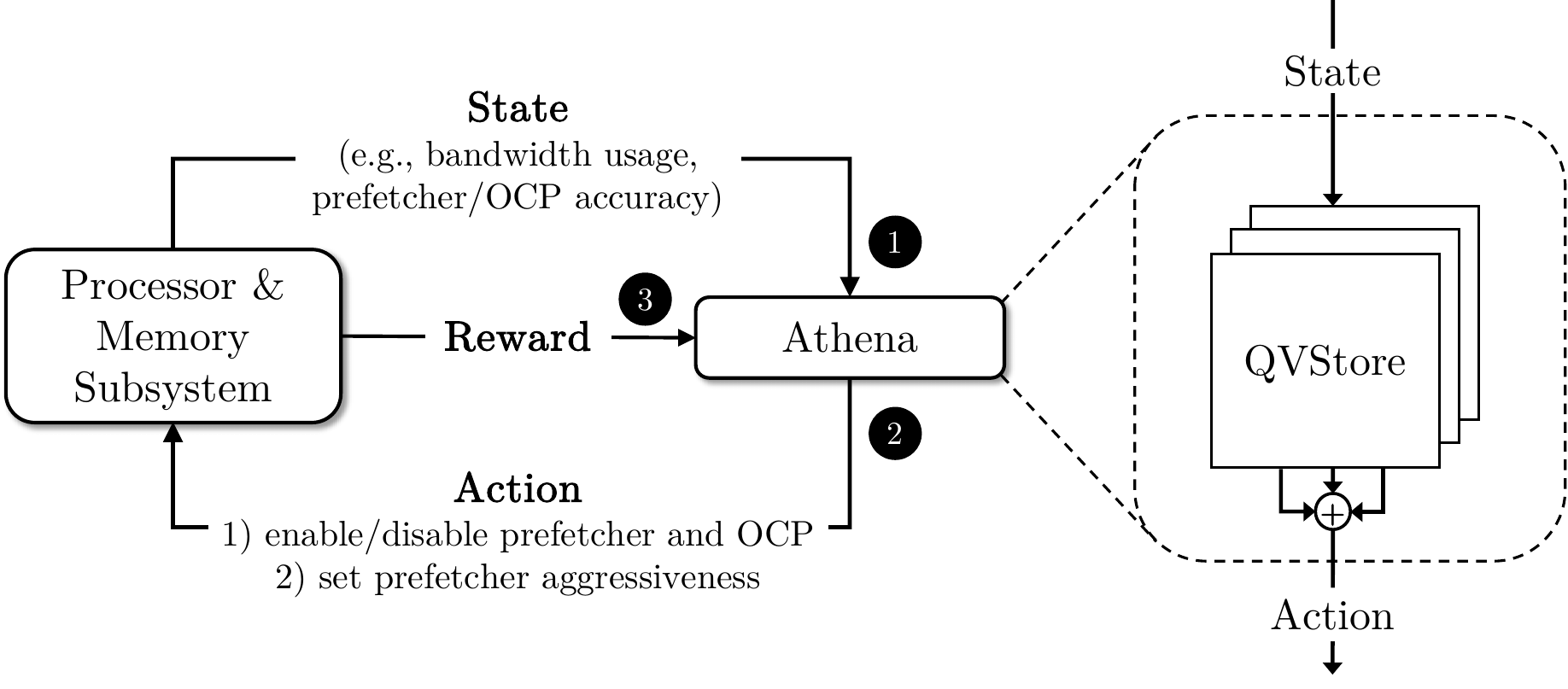}
    \caption{High-level overview of Athena as an RL agent.}
    \label{fig:athena_rl}
\end{figure}

Athena formulates the coordination of data prefetchers and the off-chip predictor as an RL problem, as illustrated in~\Cref{fig:athena_rl}. 
Here, Athena acts as an RL agent that continuously learns and adapts its prefetcher-OCP coordination policy by interacting with the processor and memory system.
As~\Cref{fig:athena_rl} shows, Athena comprises a key hardware structure, \emph{Q-Value Storage (QVStore)}, whose purpose is to store the Q-values of state-action pairs encountered during Athena's online operation.

Each timestep for Athena corresponds to a fixed-length epoch of workload execution (e.g., $N$ retired instructions). 
During an execution epoch, Athena observes and records various system-level features (e.g., prefetcher/OCP accuracy, memory bandwidth usage).
At the end of every epoch, Athena uses the recorded feature values as \emph{\textbf{state}} information to index into the QVStore (step \circled{1} in~\Cref{fig:athena_rl}) to select a coordination \emph{\textbf{action}}, i.e., whether to enable only prefetcher, only OCP, both mechanisms, or none of them (step \circled{2}).
If Athena decides to enable the prefetcher, it further determines the prefetcher aggressiveness based on the magnitude of the selected action's Q-value (see~\Cref{subsec:action}). 
At the end of every epoch, Athena receives a numerical \emph{\textbf{reward}} that measures the change in multiple system-level metrics (e.g., the number of cycles taken to execute an epoch) to evaluate the impact of its actions on system performance (step \circled{3}). 
Athena uses this reward to autonomously and continuously learn a prefetcher-OCP coordination policy, that can adapt to diverse workloads and system configurations. 

\subsection{State} \label{subsec:state}
We define the state as a vector of system-level features, where each feature encapsulates a distinct behavior of the memory subsystem (e.g., the accuracy of the prefetcher, the main memory bandwidth usage) observed during the current execution epoch.
Collectively, these features represent the runtime conditions relevant to making a prefetcher-OCP coordination decision.
While Athena can, in principle, learn a coordination policy using \emph{any} arbitrary set of features, increasing the state dimension rapidly increases the storage required to maintain the Q-values of all observed state-action pairs. 
To bound the storage overhead of Athena, we fix the state representation offline using a two-step process.
First, guided by domain knowledge, we identify a set of seven candidate system-level features that are expected to influence the effectiveness of prefetcher-OCP coordination.
\Cref{tab:feature_list} summarizes each feature, its computation method, and the rationale behind its inclusion.
Second, we perform offline feature selection using automated design-space exploration (as described in~\Cref{subsec:automated}) to determine the final subset of features that Athena uses to construct the state vector.

\begin{table}[!ht]
    \centering
    \caption{Candidate features for Athena's state representation.}
    \label{tab:feature_list}
    \footnotesize 
    \begin{tabular}{L{5em}||C{14em}|L{7.2em}}
        \thickhline
        \textbf{Feature} & \textbf{Measurement} & \textbf{Rationale} \\ 
        \thickhline
        \Tabval{\textbf{Prefetcher accuracy}} & 
        \Tstrut
        \begin{minipage}{14em}
            \vskip 6pt
            \centering
            $\dfrac{\text{\# demand hits}}{\text{\# prefetches issued}}$
        \end{minipage}
        \Bstrut & \Tabval{Effectiveness of prefetching} \\
        \hline 
        \Tabval{\textbf{OCP accuracy}} & 
        \Tstrut
        \begin{minipage}{14em}
            \vskip 6pt
            \centering
            $\dfrac{\text{\# off-chip demand hits}}{\text{\# off-chip predictions}}$
        \end{minipage}
        \Bstrut & \Tabval{Effectiveness of OCP} \\
        \hline
        \Tabval{\textbf{Bandwidth usage}} & 
        \Tstrut
        \begin{minipage}{14em}
            \vskip 6pt
            \centering
            $\dfrac{\text{used main memory bandwidth}}{\text{peak main memory bandwidth}}$
        \end{minipage}
        \Bstrut & \Tabval{Memory bus pressure} \\
        \hline
        \Tabval{\textbf{Cache pollution}} & 
        \Tstrut
        \begin{minipage}{14em}
            \vskip 6pt
            \centering
            $\dfrac{\text{\# prefetch-evicted demand misses }}{\text{\# total demand misses}}$
        \end{minipage}
        \Bstrut & \Tabval{Interference by prefetcher} \\
        \hline 
        \Tabval{\textbf{Prefetch bandwidth}} & 
        \Tstrut
        \begin{minipage}{14em}
            \vskip 6pt
            \centering
            $\dfrac{\text{\# prefetch requests to DRAM}}{\text{\# total DRAM requests}}$
        \end{minipage}
        \Bstrut & \Tabval{Prefetcher's share of memory traffic} \\
        \hline
        \Tabval{\textbf{OCP bandwidth}} & 
        \Tstrut
        \begin{minipage}{14em}
            \vskip 6pt
            \centering
            $\dfrac{\text{\# OCP requests to DRAM}}{\text{\# total DRAM requests}}$
        \end{minipage}
        \Bstrut & \Tabval{OCP's share of memory traffic} \\
        \hline
        \Tabval{\textbf{Demand bandwidth}} & 
        \Tstrut
        \begin{minipage}{14em}
            \vskip 6pt
            \centering
            $\dfrac{\text{\# demand requests to DRAM}}{\text{\# total DRAM requests}}$
        \end{minipage}
        \Bstrut & \Tabval{Demand's share of memory traffic} \\
        \thickhline
    \end{tabular}
\end{table}

\subsection{Action} \label{subsec:action}

Athena's action space consists of four coordination decisions, i.e., whether to enable (1) only prefetcher, (2) only OCP, (3) both mechanisms, or (4) none of them. 
These actions allow Athena to explicitly coordinate the prefetcher and the OCP at a coarse granularity, i.e., only enabling or disabling any given mechanism as a whole.
However, when Athena selects an action that enables prefetching, it further determines the prefetcher aggressiveness.
This aggressiveness is derived directly from the learned Q-values as a function of the \emph{relative difference} between the Q-value of the selected action and the average Q-value of the remaining actions.
The underlying rationale for this Q-value-driven prefetcher aggressiveness control is that the magnitude of the selected action's Q-value implicitly encodes Athena's confidence in the prefetcher's effectiveness.
A larger separation between the Q-value of the selected action (that enables the prefetcher, either without or with the OCP) and those of the alternative actions indicates stronger historical evidence that enabling prefetching is beneficial in the current state, thus warranting more aggressive prefetching.
On the other hand, a smaller separation reflects uncertainty in prefetcher effectiveness, prompting more conservative prefetching.
\Cref{algo:pref_deg} formalizes this Q-value-based prefetcher aggressiveness control mechanism.\footnote{Here, we represent prefetcher aggressiveness using the prefetch degree, i.e., the number of prefetch requests issued per demand trigger. However, the proposed Q-value-driven aggressiveness control mechanism can also be applied to alternative aggressiveness definitions.}

\begin{algorithm}[!ht]
\footnotesize
\begin{algorithmic}[1]
\Procedure{SelectPrefetchDegree}{}
    \State $a^* \gets \arg\max_{a \in \mathcal{A}} Q(a)$
    \State $avg \gets$ average Q-value of all remaining actions
    \State $\Delta Q \gets Q(a^*) - avg$ \Comment{compute Q-value confidence}
    \State $r \gets \min(1,\, \Delta Q / \tau)$ \Comment{normalize confidence w.r.t. hyperparam $\tau$}
    \State $d \gets \lfloor r \cdot d_{\max} \rfloor$ \Comment{adjust prefetch degree}
    \State \Return $d$
\EndProcedure
\end{algorithmic}
\caption{Q-value-driven prefetcher aggressiveness control}
\label{algo:pref_deg}
\end{algorithm}

First, Athena selects the action $a^*$ with the highest Q-value.
It then estimates the confidence of this decision by comparing $Q(a^*)$ against the average Q-value of all remaining actions.
The resulting confidence ratio $\Delta Q$ captures how strongly the selected action is preferred over the alternatives.
Using this confidence signal, Athena determines the final prefetch degree as a fraction of $d_{\max}$, where $d_{\max}$ denotes the number of prefetch requests that the underlying prefetcher can issue while operating at full aggressiveness.
If this confidence ratio exceeds a hyperparameter $\tau$, Athena enables the prefetcher at full aggressiveness, reflecting high confidence in the benefits of prefetching in the current system state.
Otherwise, Athena scales the prefetch degree proportionally to $\Delta  / \tau$, issuing fewer prefetch requests.

\subsection{Reward} \label{subsec:reward}
The reward defines the optimization objective for Athena.
Prior works often use the change in instructions committed per cycle (IPC) as the \emph{only} system-level metric to train the RL agent~\cite{mab, jalili2022managing, micro_mama, eris2022puppeteer}.
However, a change in IPC may originate from two different sources: 
(1) the coordination actions taken by the agent, and 
(2) the inherent variations in workload behavior, that are independent of the agent's actions. 
As such, using IPC as the sole reward can be unreliable and may mislead the learned policy.

To address this limitation, Athena introduces a composite reward framework that explicitly separates the effects of Athena's action on the system from the inherent variations in workload behavior.
More specifically, we define the reward for Athena at a given timestep $t$ using two components:
(1) \emph{correlated reward} ($R_{t}^{corr}$), which encapsulates the effect of Athena's action on the system,
and (2) \emph{uncorrelated reward} ($R_{t}^{uncorr}$), which encapsulates the inherent change in program behavior.
The overall reward ($R_t$) is defined using these two component rewards as:
\begin{equation} 
    \label{eq:reward}
    R_t = R_{t}^{corr} - R_{t}^{uncorr}
\end{equation}

By subtracting the uncorrelated reward component from the correlated reward, Athena aims to isolate the performance impact that is causally attributable to its coordination actions from variations induced by inherent workload behavior.
This allows Athena to learn a higher-performing prefetcher-OCP coordination policy than it would have otherwise learned using a single, conflated reward signal (as we show in~\Cref{subsec:eval_ablation}).

\paraheading{Correlated Reward.}
We define the correlated reward at a given timestep $t$ as a \emph{linear combination} of the changes in the constituent system-level metrics that are influenced by Athena's actions in two consecutive timesteps. 
Formally, the correlated reward, $R_{t}^{corr}$, is defined as:

\begin{equation} 
    \label{eq:corr_reward}
    R_{t}^{corr} = \sum_{i} \lambda_i \cdot \Delta M_{i,t}^{corr}
\end{equation}

Here, $\Delta M_{i,t}^{corr}$ denotes the change in the $i$-th correlated system-level metric observed between timesteps $(t-1)$ and $t$, and $\lambda_i$ is a hyperparameter that captures the relative weight of this metric to the overall reward.\footnote{The value of each weight parameter is tuned offline using the automated design-space exploration described in~\Cref{subsub:hyp_tuning}.}

In principle, any system-level metric that is directly affected by Athena's actions (e.g., execution cycles, number of last-level cache misses) can be incorporated into the correlated reward.
In practice, we conduct an offline sensitivity analysis over a broad set of candidate metrics and select three metrics shown in~\Cref{tab:reward_decomp} as the constituents of the correlated reward that provide a stable and informative learning signal across diverse workloads and system configurations.\footnote{The extended version of this paper~\cite{athena_extended} reports the complete list of system-level metrics evaluated during the offline sensitivity analysis.}

\paraheading{Uncorrelated Reward.}
We define the uncorrelated reward at a given timestep $t$ as a \emph{linear combination} of the changes in its constituent metrics that are largely independent of Athena's actions but are influenced by the inherent variations in workload behavior. 
Formally, the uncorrelated reward, $R_{t}^{uncorr}$, is defined as:

\begin{equation} 
    \label{eq:uncorr_reward}
    R_{t}^{uncorr} = \sum_{j} \lambda_j \cdot \Delta M_{j,t}^{uncorr}
\end{equation}

Here, $\Delta M_{j,t}^{uncorr}$ denotes the change in the $j$-th uncorrelated metric observed between timesteps $(t-1)$ and $t$, and $\lambda_j$ is a hyperparameter that captures its relative weight to the overall reward.
While any metric that is affected by changes in workload behavior can be incorporated into the uncorrelated reward, we select two metrics shown in~\Cref{tab:reward_decomp} as the constituents of the uncorrelated reward, based on offline sensitivity analysis.

\begin{table}[h]
    \centering
    \caption{Constituent metrics of Athena's reward.}
    \label{tab:reward_decomp}
    \small
    \begin{tabular}{l|l|c}
        \thickhline
        & \textbf{Constituent metric} & \textbf{Weight} \\ 
        \thickhline
        \multirow{3}{*}{\textbf{\( R_{t}^{\text{corr}} \)}} 
        & \# Cycles & $\lambda_{\text{cycle}} $ \\
        & \# LLC misses & $\lambda_{\text{LLC}_\text{m}}$ \\
        & LLC miss latency & $\lambda_{\text{LLC}_\text{t}}$ \\ 
        \hline
        \multirow{2}{*}{\textbf{\( R_{t}^{\text{uncorr}} \)}} 
        & \# Load instructions & $\lambda_{\text{load}}$ \\ 
        & \# Mispredicted branches &$\lambda_{\text{MBr}}$ \\
        \thickhline
    \end{tabular}
\end{table}

\section{Athena: Detailed Design}\label{sec:implementation}

\subsection{QVStore Organization}\label{subsec:qvstore}
The QVStore maintains the Q-values of all state-action pairs encountered by Athena during online execution. 
Unlike prior RL-based approaches that rely on deep neural networks to approximate Q-values~\cite{jalili2022managing}, or operate without state information~\cite{mab, micro_mama}, Athena adopts a lightweight and hardware-friendly tabular organization for storing Q-values, tailored for low-latency access and online updates.

\begin{figure}[!ht] 
\centering \includegraphics[width=\columnwidth]{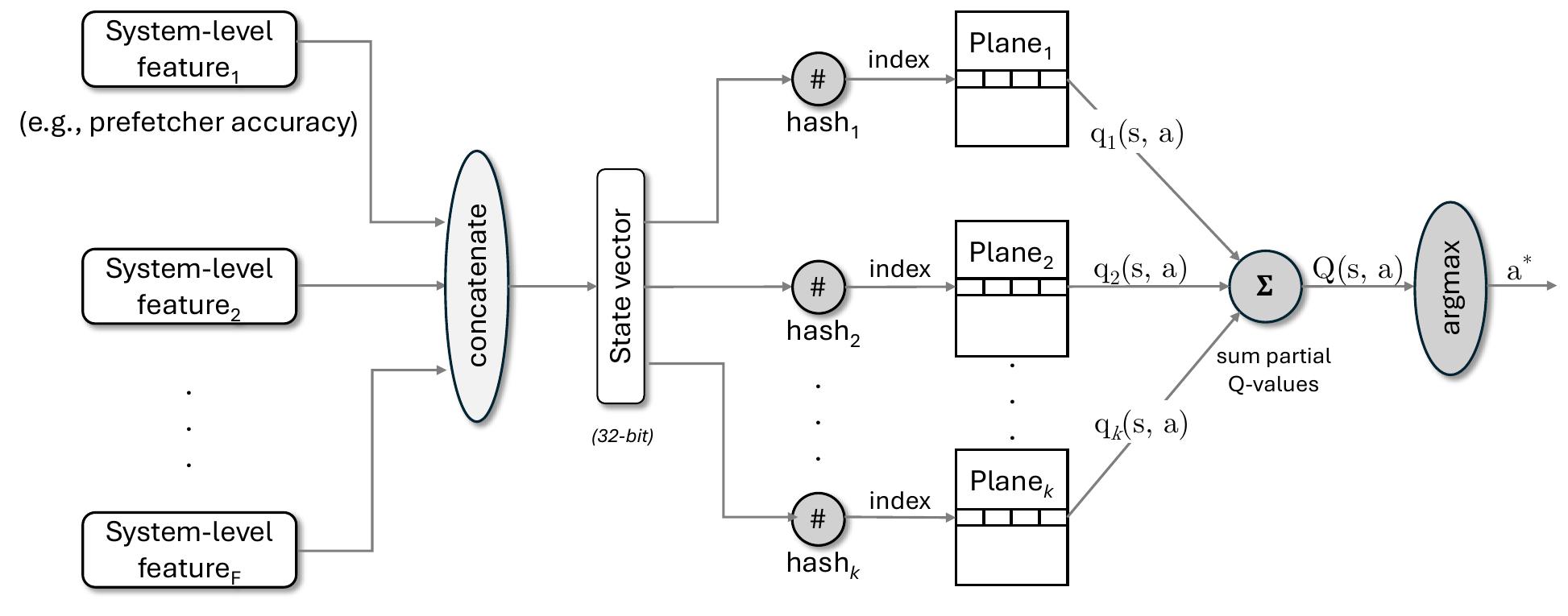}
\caption{Organization of QVStore.} 
\label{fig:at_mechanism} 
\end{figure}

\Cref{fig:at_mechanism} demonstrates the organization of QVStore and the procedure for retrieving the Q-value of a given state $S$ and action $A$.
As the number of Q-values that need to be stored for every possible state-action pair grows rapidly with 
(1) the number of constituent features in a state vector, 
and (2) the number of bits used to represent each feature, naively implementing the QVStore as a monolithic table quickly becomes impractical due to its storage overhead.
Such a design also incurs prohibitive access latency and power overhead, making it unsuitable for designing a timing-critical hardware RL agent.

To address these challenges, Athena organizes the QVStore as a partitioned structure comprising $k$ independent tables, each of which we call a \emph{plane}. 
Each plane stores a partial Q-value of a given state-action pair. 
This partitioned organization enables Athena to decouple the storage cost from the full combinatorial state space while simultaneously supporting fast, parallel access.

For a given state-action pair, Athena retrieves its corresponding Q-value from the partitioned QVStore in three stages, as shown in~\Cref{fig:at_mechanism}.
First, Athena constructs the state vector by concatenating all feature values.
Second, Athena applies $k$ distinct hash functions to the state vector, each producing an index to a plane to retrieve the corresponding partial Q-value \emph{in parallel}.
Third, Athena computes the final Q-value by summing the partial Q-values across all planes.
At the end of each epoch, Athena updates the Q-value using the SARSA update rule (see~\Cref{sec:background}), applying the update independently to each plane.

The partitioned, multi-hash organization of QVStore provides two key benefits. 
First, hashing the same state into multiple planes strikes a balance between generalization and resolution: similar states are likely to collide in at least some planes, promoting value sharing, while dissimilar states are likely to be de-aliased through independent hashes.
Second, partitioning keeps each plane compact, enabling low-latency, energy-efficient parallel reads and updates.
Together, these properties substantially reduce storage overhead while supporting fast and scalable Q-value access and updates during Athena's operation.

\subsection{State Measurement}\label{subsec:state_measurement}

\subsubsection{\textbf{Prefetcher Accuracy}}
Athena employs a Bloom filter~\cite{bloom} to track prefetcher accuracy.
For every prefetch issued, the corresponding address is inserted into the Bloom filter.  
Upon each demand access, Athena queries the filter to determine whether the address was prefetched.  
Prefetch accuracy is computed as the ratio of demand accesses that hit in the filter over the number of issued prefetches. 
Athena resets the filter at the end of each epoch.

\subsubsection{\textbf{OCP Accuracy}}
Athena measures OCP accuracy using two simple counters.
When OCP predicts that a demand load will go off-chip, it issues a speculative request to the memory controller to start fetching the data directly from main memory. 
Athena tracks such predictions using a dedicated counter.
When a demand request misses all cache levels and arrives at the memory controller, it indicates that this request indeed went off-chip.
Athena increments another counter to track correct predictions.
OCP accuracy is computed as the ratio of correctly predicted off-chip accesses over the total number of off-chip predictions. 
Athena resets both counters at the end of each epoch.

\subsubsection{\textbf{Prefetch-Induced Cache Pollution}}
Athena uses a Bloom filter~\cite{bloom} to track prefetch-induced cache pollution at the last-level cache (LLC).
When a cache block is evicted from the LLC for a prefetch fill, Athena inserts the evicted address into the filter.  
If the address of a subsequent LLC miss hits the filter, Athena increments a dedicated counter.
Athena resets the filter and the counter at the end of each epoch. 
This method of measuring prefetch-induced cache pollution is similar to prior works~\cite{fdp, ebrahimi2009techniques}.

\subsection{Automated Design-Space Exploration} \label{subsec:automated}
We employ automated design-space exploration (DSE) to select the optimal state features (see \Cref{subsec:state}), reward weights (see~\Cref{subsec:reward}), and hyperparameters ($\alpha$, $\gamma$, $\epsilon$, $\tau$, and epoch length).
To prevent overfitting, DSE is conducted on $20$ selected workloads, which are \emph{not included} in the final set of $100$ workloads.
We perform DSE using Cache Design~1 (CD1) in a single-core configuration, with POPET as the OCP and Pythia as the L2C prefetcher (see~\Cref{sec:meth_sys_config}).
The configuration that achieves the best performance across the $20$ selected workloads is then applied \textit{unaltered} to the full set of $100$ evaluation workloads, to all other cache designs (see~\Cref{sec:eval_sc}), OCPs (see~\Cref{sec:eval_sen_cd1}), prefetchers (see~\Cref{sec:eval_sen_cd4}), and multi-core experiments (see~\Cref{sec:eval_mc}).

\subsubsection{\textbf{Feature Selection}}
We derive the program features through an iterative process. 
Starting from the initial set of seven candidate features (see \cref{subsec:state}), we begin with the feature that yields the highest standalone performance gain. 
In each iteration, we include the feature that provides the greatest additional performance improvement while retaining the previously selected features.
We observe diminishing performance gains after the fourth iteration. 
Consequently, we fix the feature set to the following four: prefetcher accuracy, OCP accuracy, bandwidth usage, and prefetch-induced cache pollution, as summarized in \Cref{table:config}.\\

\begin{table}[!ht]
  \centering
  \caption{Final Athena configuration derived through automated design-space exploration.}
  \label{table:config}
  \footnotesize
  \begin{tabular}{L{8em}||L{17em}}
    \thickhline
    \Tabval{\textbf{Category}} & \Tabval{\textbf{Final Values}} \\
    \thickhline
    \Tabval{\textbf{\emph{Selected Features}}} & \Tabval{(1) prefetcher accuracy, (2) OCP accuracy, (3) bandwidth usage, (4) prefetch-induced cache pollution} \\
    \hline
    \Tabval{\textbf{\emph{Reward Weights}}} & \Tabval{$\lambda_{\text{cycle}} = 1.6$, $\lambda_{\text{LLC}_\text{m}} = 0.0$, $\lambda_{\text{LLC}_\text{t}} = 0.0$, $\lambda_{\text{load}} = 0.6$, $\lambda_{\text{MBr}} = 1.0$} \\
    \hline
    \Tabval{\textbf{\emph{Hyperparameters}}} & \Tabval{$\alpha = 0.6$, $\gamma = 0.6$, $\epsilon = 0.0$, $\tau = 0.12$, Epoch length ($N$) = $2K$ instructions} \\
    \thickhline
  \end{tabular}
\end{table}

\subsubsection{\textbf{Reward and Hyperparameter Tuning}} \label{subsub:hyp_tuning}
We use grid search~\cite{grid_search1,grid_search2} to tune reward weights (see~\Cref{subsec:reward}) and hyperparameters (see~\Cref{subsec:bg_rl}). 
For each reward weight and hyperparameter, we define a search range and discretize it into equally spaced grid points.
The learning rate \(\alpha\), the discount factor \(\gamma\), and the exploration rate \(\epsilon\) are searched over $[0, 1]$ in steps of $0.1$. 
Reward weights \(\lambda_{i}\) and \(\lambda_{j}\) are searched over $[0, 2]$ in steps of $0.2$.
The final reward weights and hyperparameters are shown in \Cref{table:config}.

\subsection{Overhead Analysis}\label{subsec:storageoverhead}

\subsubsection{\textbf{Storage Overhead}} \label{subsec:overhead_storage}
\Cref{table:overhead} summarizes the storage overhead of Athena.
The QVStore is organized into eight planes, each containing $64$ rows and $4$ columns (one per action). Each entry stores an $8$-bit Q-value. 
To size the Bloom filter used for prefetcher accuracy tracking, we experimentally observe an average of $49$ prefetch requests per epoch (i.e., $2K$ retired instructions), with a standard deviation (SD) of $50$. 
We therefore size the Bloom filter for prefetcher accuracy tracking at $4096$ bits, which yields a false positive rate of $1\%$ when accommodating three SDs above the average (i.e., $199$ requests).
Similarly, we observe an average of $62$ LLC evictions per epoch, with an SD of $58$. 
Thus, we size the Bloom filter for prefetch-induced cache pollution tracking at $4096$ bits, providing a false positive rate of $1\%$ when inserting three SDs more evictions than the average (i.e., $236$ evictions).
\\

\begin{table}[!ht]
  \centering
  \caption{Storage overhead of Athena.}
  \label{table:overhead}
  \footnotesize
  \begin{tabular}{l||l||r}
    \thickhline
    \textbf{Structure} & \textbf{Description} & \textbf{Size} \\
    \thickhline
    \multirow{2}[1]{*}{\textbf{QVStore}} & \# planes = $8$,  \# rows = $64$ & \multirow{2}[1]{*}{{2~KB}} \\
    & \# columns = $4$, entry size = $8$ bits & \\
    \hline
    \textbf{Accuracy Tracker} & $4096$-bit Bloom filter, $2$ hashes & {0.5~KB} \\
    \hline
    \textbf{Pollution Tracker} & $4096$-bit Bloom filter, $2$ hashes & {0.5~KB} \\
    \thickhline
    \textbf{Total} & & \textbf{3~KB} \\
    \thickhline
  \end{tabular}
\end{table}

\subsubsection{\textbf{Latency Overhead}} \label{subsec:overhead_latency}
At the end of every epoch, Athena needs to compute the overall reward from its constituent partial rewards and update the QVStore, both of which incur considerable computation overhead. 
To account for such computations, we model Athena with a delayed QVStore update latency of $50$ cycles (i.e., the QVStore is updated $50$ cycles after the end of an epoch).
We also sweep the update latency and observe that Athena's performance benefit is not sensitive to the update latency. 
This is because Athena needs to query the updated QVStore only at the end of the current epoch (i.e., $2K$ retired instructions), which takes considerably longer than even the pessimistic estimate of the update latency.
\section{Methodology} 
\label{sec:methodology}

We evaluate Athena using the ChampSim trace-driven simulator~\cite{champsim}.
We faithfully model an Intel Golden Cove-like microarchitecture~\cite{goldencove}, including its large reorder buffer (ROB), multi-level cache hierarchy, and publicly reported on-chip cache access latencies~\cite{llc_lat1, llc_lat2, l3_lat_compare1}. 
Table~\ref{table:sim_params} summarizes the key microarchitectural parameters.
The source code of Athena is freely available at~\cite{athena_github}.
\\

\begin{table}[!ht]
    \centering
    \caption{Simulated system parameters.}
    \label{table:sim_params}
    \footnotesize
    \begin{tabular}{L{4.5em}||L{23.2em}}
        \thickhline
        \hline
        \Tabval{\textbf{Core}} & \Tabval{$6$-wide fetch/issue/commit, $512$-entry ROB, $128$-entry LQ, $72$-entry SQ, perceptron branch predictor~\cite{perceptron}, $17$-cycle misprediction penalty} \\
        \hline
        \Tabval{\textbf{L1I/D}} & \Tabval{Private, $48$KB, $64$B line, $12$-way, $16$ MSHRs, LRU, $4/5$-cycle round-trip latency} \\
        \hline
        \Tabval{\textbf{L2C}} & \Tabval{Private, $1.25$MB, $64$B line, $20$-way, $48$ MSHRs, LRU, $15$-cycle round-trip latency~\cite{llc_lat2}} \\
        \hline
        \Tabval{\textbf{LLC}} &  \Tabval{Shared, $3$MB/core, $64$B line, $12$-way, $64$ MSHRs/slice, SHiP~\cite{ship}, $55$-cycle round-trip latency~\cite{llc_lat1, llc_lat2}} \\
        \hline
        \Tabval{\textbf{Main Memory}} & \Tabval{$1$ rank per channel, $8$ banks per rank, $64$-bit data bus, $2$~KB row buffer, $t_{\text{RCD}}$ = $t_{\text{RP}}$ = $t_{\text{CAS}}$ = $12.5$~ns; DDR4 DRAM with $3.2$ GB/s per core}\\
        \thickhline
    \end{tabular}
\end{table}

\subsection{Workloads} \label{sec:meth_workloads}

We evaluate Athena using a diverse set of workload traces spanning \texttt{SPEC CPU 2006}~\cite{spec2006}, \texttt{SPEC CPU 2017}~\cite{spec2017}, \texttt{PARSEC}~\cite{parsec}, \texttt{Ligra}~\cite{ligra}, and real-world commercial workloads from the first Value Prediction Championship (\texttt{CVP}~\cite{cvp1}). 
We only consider workloads in our evaluation that have at least $3$ LLC misses per kilo instructions (MPKI) in the no-prefetching and no-OCP system.
In total, we use $100$ memory-intensive single-core workload traces, as summarized in Table~\ref{table:workloads}.
\texttt{SPEC CPU 2006} and \texttt{SPEC CPU 2017} workloads are collectively referred to as \texttt{SPEC}. All the workload traces used in our evaluation are freely available online~\cite{athena_github}.
\\

\begin{table}[htbp]
  \centering
  \caption{Workloads used for evaluation.}
  \label{table:workloads}%
  \footnotesize
    \begin{tabular}{L{6.6em}C{3.9em}L{16.2em}}
    \toprule
    \Tabval{\textbf{Suite}} & \Tabval{\textbf{\# Traces}} & \Tabval{\textbf{Example Workloads}} \\
    \toprule
    \Tabval{SPEC CPU 2006} & \Tabval{29} & \Tabval{astar, gobmk, GemsFDTD, leslie3d, libquantum, milc, omnetpp, sphinx3, \dots} \\ 
    \Tabval{SPEC CPU 2017} & \Tabval{20} & \Tabval{bwaves, cactuBSSN, cam4, fotonik3d, gcc, lbm, mcf, xalancbmk, \dots} \\
    \Tabval{PARSEC} & \Tabval{13} & \Tabval{canneal, facesim, fluidanimate, raytrace, streamcluster, \dots} \\
    \Tabval{Ligra} & \Tabval{13} & \Tabval{BC, BFS, BFSCC, CF, PageRank, PageRankDelta, Radii, Triangle, \dots} \\
    \Tabval{CVP} & \Tabval{25}  & \Tabval{integer, floating point, \dots} \\
    \bottomrule
    \end{tabular}
\end{table}

For multi-core evaluation, we construct $90$ four-core and $90$ eight-core workload mixes, each comprising three categories.
(1) $30$ \emph{prefetcher-adverse mixes}: each workload is randomly selected from the prefetcher-adverse workloads.
(2) $30$ \emph{prefetcher-friendly mixes}: each workload is randomly selected from the prefetcher-friendly workloads.
(3) $30$ \emph{random mixes}: workloads are drawn uniformly at random from the entire set of $100$ workloads.

For all single-core simulations, we perform a warm-up of $100$ million (M) instructions. 
\texttt{SPEC} workloads are simulated for $500$M, \texttt{PARSEC}, \texttt{Ligra}, and \texttt{CVP} workloads for $150$M instructions.
For multi-core simulations, each core performs a warm-up of $10$M instructions, followed by simulating $50$M instructions.
The workloads are replayed as needed to ensure all cores reach the required number of simulated instructions.

\subsection{Evaluated Prior Prefetcher Control Policies} 
\label{sec:meth_sys_config_pol}

We compare Athena against three prior prefetcher control policies: 
TLP~\cite{jamet2024tlp}, HPAC~\cite{ebrahimi2009coordinated}, and MAB~\cite{mab}.

\subsubsection{\textbf{Two Level Perceptron}}
TLP~\cite{jamet2024tlp} explicitly incorporates OCP into its coordination framework and combines it with prefetch filtering~\cite{jamet2024tlp} at the L1D. We adopt the same features, prediction thresholds ($\tau_{\text{low}}$, $\tau_{\text{high}}$), and filtering threshold ($\tau_{\text{pref}}$) as specified in~\cite{jamet2024tlp}.

\subsubsection{\textbf{Hierarchical Prefetcher Aggressiveness Control}}
HPAC~\cite{ebrahimi2009coordinated} compares various system-level feature values against static thresholds to make prefetch control decisions. Although not designed with OCP in mind, we adapt HPAC to coordinate prefetchers and OCP. We use three system-level features for local aggressiveness control: prefetcher accuracy, OCP accuracy, and main-memory bandwidth usage.
We use the bandwidth needed by each core as the feature for global aggressiveness control.
For each feature, we tune its static threshold via extensive grid search using the same set of tuning workloads that we use to tune Athena (see~\cref{subsec:automated}).

\subsubsection{\textbf{Micro-Armed Bandit}}\label{subsubsec:mab}
MAB~\cite{mab} uses a multi-armed bandit algorithm~\cite{multi_arm1,multi_arm2} for its decision-making. 
Although originally not designed for OCP, we adapt MAB to coordinate OCP with prefetchers. 
MAB selects whether to enable the prefetcher and OCP based on previous reward feedback derived from the system's IPC.
Thus, our implementation of MAB uses four (eight) arms while coordinating one OCP in the presence of one (two) prefetcher(s).
We find the best-performing hyperparameters via grid search~\cite{grid_search1,grid_search2} using the same set of tuning workloads that we use to tune Athena (see~\cref{subsec:automated}).

\subsection{Evaluated Cache Designs} \label{sec:meth_sys_config}
To demonstrate Athena's adaptability, we evaluate Athena across diverse cache designs (CDs), as summarized in \Cref{tab:design_comparison}. 
All these cache designs include an OCP alongside the three-level cache hierarchy and differ only in the number and placement of prefetchers, closely mimicking cache hierarchy designs found in commercial processors.    
For each cache design, we identify suitable prior approaches as comparison points. HPAC and MAB can be adapted to all four cache designs to coordinate OCP with prefetchers. 
In contrast, TLP, by design, is restricted to cache designs involving an L1D prefetcher (i.e., CD2 and CD4; see \cref{subsubsec:prior}).
Unless stated otherwise, CD1 serves as the default cache configuration.
\\

\begin{table}[htbp]
\centering
\caption{Evaluated cache designs (CD) and corresponding comparison points. L1D = L1 data cache; L2C = L2 cache.}
\label{tab:design_comparison}
\footnotesize
    \begin{tabular}{C{4em}ll}
    \toprule
    \Tabval{\textbf{Design\#}} & \Tabval{\textbf{Description}} & \Tabval{\textbf{Comparison Points}} \\
    \toprule
    \Tabval{\textbf{CD1}} & \Tabval{OCP + 1 L2C prefetcher} & \Tabval{HPAC, MAB} \\
    \Tabval{\textbf{CD2}} & \Tabval{OCP + 1 L1D prefetcher} & \Tabval{HPAC, MAB, TLP} \\
    \Tabval{\textbf{CD3}} & \Tabval{OCP + 2 L2C prefetchers} & \Tabval{HPAC, MAB} \\
    \Tabval{\textbf{CD4}} & \Tabval{OCP + 1 L1D + 1 L2C prefetcher} & \Tabval{HPAC, MAB, TLP} \\
    \bottomrule
    \end{tabular}
\end{table}

\subsection{Evaluated Data Prefetchers} \label{sec:meth_sys_config_pref}

We assess Athena's flexibility by integrating six prefetchers, i.e., IPCP~\cite{ipcp}, Berti~\cite{navarro2022berti}, Pythia~\cite{pythia}, SPP~\cite{spp} with perceptron-based prefetch filter (PPF)~\cite{ppf}, SMS~\cite{sms}, and MLOP~\cite{mlop}, at various levels of the cache hierarchy.
IPCP and Berti are evaluated at L1D and are trained using all memory requests looking up the L1D. 
Pythia, SPP+PPF, MLOP, and SMS operate at L2C and are trained using all memory requests looking up the L2C. 
All prefetchers prefetch in the physical address space.
Unless stated otherwise, we use IPCP as the default L1D prefetcher, and Pythia as the default L2C prefetcher.

\subsection{Evaluated Off-Chip Predictors} \label{sec:meth_sys_config_ocp}

We also evaluate Athena across three OCPs: POPET~\cite{hermes}, HMP~\cite{yoaz1999speculation}, and TTP~\cite{lp, hermes}.
POPET uses a hashed-perceptron network with five program features to make accurate off-chip predictions. We evaluate the exact POPET configuration presented in~\cite{hermes}.
HMP combines three prediction techniques analogous to hybrid branch prediction: local~\cite{yeh1991two, yeh1992alternative}, gshare~\cite{mcfarling1993combining}, and gskew~\cite{michaud1997trading}. 
TTP, as introduced by~\cite{lp, hermes}, predicts off-chip loads by tracking cacheline tags. 
We evaluate the exact TTP configuration open-sourced by~\cite{hermes}.
Similar to prior work~\cite{hermes}, all speculative load requests issued by the OCPs incur a $6$-cycle latency before reaching the memory controller.
Unless stated otherwise, we use POPET as the default OCP.
Table~\ref{table:eval_sys} summarizes the storage overhead of all evaluated mechanisms.

\begin{table}[htbp]
  \centering
  \caption{Storage overhead of all evaluated mechanisms.}
  \label{table:eval_sys}
  \footnotesize
    \begin{tabular}{|L{1em}||L{0.8em}|L{18em}||R{4em}|}
    \hline
    \multirow{6}[2]{*}{\begin{sideways}\textbf{Prefetchers}\end{sideways}} 
      & \multirow{2}[1]{*}{\begin{sideways}\textbf{L1D}\end{sideways}} 
          & \Tstrut \textbf{IPCP}~\cite{ipcp}, as an L1D-only prefetcher & \textbf{0.7 KB} \\ \cline{3-4}
      &   & \Tstrut \textbf{Berti}, as configured in~\cite{navarro2022berti} & \textbf{2.55 KB} \\ \cline{2-4}
      & \multirow{4}[2]{*}{\begin{sideways}\textbf{L2C}\end{sideways}}
          & \Tstrut \textbf{Pythia}, as configured in~\cite{pythia} & \textbf{25.5 KB} \\ \cline{3-4}
      &   & \Tstrut \textbf{SPP+PPF}, as configured in~\cite{spp, ppf} & \textbf{39.3 KB} \\ \cline{3-4}
      &   & \Tstrut \textbf{MLOP}, as configured in~\cite{mlop} & \textbf{8 KB} \\ \cline{3-4}
      &   & \Tstrut \textbf{SMS}, as configured in~\cite{sms} & \textbf{20 KB} \\
    \hline
    \addlinespace[0.5em]
    \hline
    \multirow{3}[2]{*}{\begin{sideways}\textbf{OCPs}\end{sideways}} 
      & \multicolumn{2}{l||}{\Tstrut \textbf{POPET}~\cite{hermes}, with 5 features} & \textbf{4 KB} \\ \cline{2-4}
      & \multicolumn{2}{l||}{\Tstrut \textbf{HMP}~\cite{yoaz1999speculation}, with 3 component predictors} & \textbf{11 KB} \\ \cline{2-4}
      & \multicolumn{2}{l||}{\Tstrut \textbf{TTP}~\cite{lp}, with metadata budget \textasciitilde L2 cache size} & \textbf{1536 KB} \\
    \hline
    \addlinespace[0.5em]
    \hline
    \multirow{4}[1]{*}{\begin{sideways}\textbf{Policies}\end{sideways}} 
      & \multicolumn{2}{l||}{\Tstrut \textbf{TLP}, as in~\cite{jamet2024tlp}} & \textbf{6.98 KB} \\ \cline{2-4}
      & \multicolumn{2}{l||}{\Tstrut \textbf{HPAC}~\cite{ebrahimi2009coordinated}, adapted for OCP} & \textbf{0.5 KB} \\ \cline{2-4}
      & \multicolumn{2}{l||}{\Tstrut \textbf{MAB}~\cite{mab}, adapted for OCP} & \textbf{0.1 KB} \\ \cline{2-4}
      & \multicolumn{2}{l||}{\Tstrut \emph{\textbf{Athena (this work)}}} & \textbf{3 KB} \\
    \hline
    \end{tabular}
\end{table}
\section{Evaluation} \label{sec:evaluation}

\subsection{Single-Core Evaluation Overview} \label{sec:eval_sc}

\subsubsection{\textbf{CD1: OCP with One L2C Prefetcher}}\label{subsubsec:L2C}
\begin{figure}[!ht]
    \centering
    \includegraphics[width=\columnwidth]{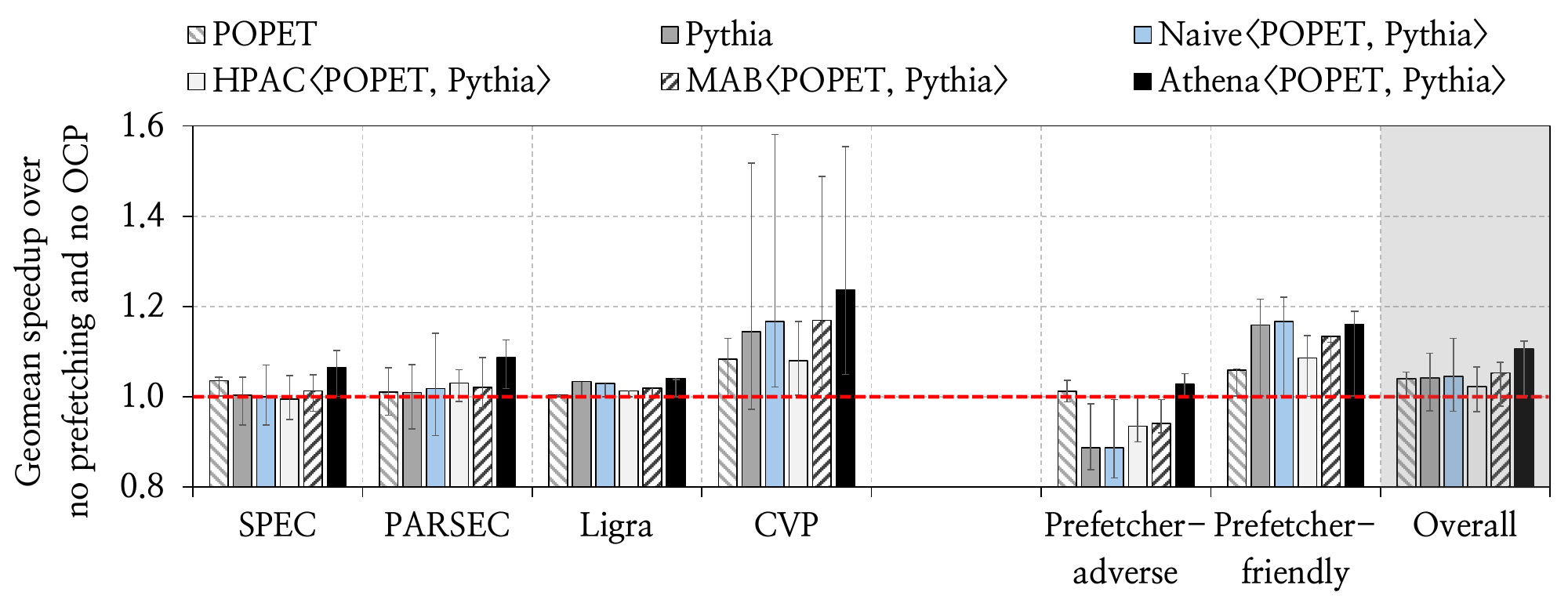}
    \caption{Speedup in cache design 1 (CD1).}
\label{fig:cd1}
\end{figure}

\Cref{fig:cd1} shows the performance improvement of Naive, HPAC, MAB, and Athena when coordinating POPET as the OCP and Pythia as the L2C prefetcher.
We make two key observations.
First, in prefetcher-adverse workloads, Naive degrades performance by $11.1\%$ compared to the baseline with no prefetching or OCP.
This degradation arises because Pythia negatively impacts performance in these workloads, undermining POPET's gains. 
In contrast, Athena dynamically identifies that POPET is beneficial and improves performance by $14.0\%$ over Naive, even surpassing POPET's standalone performance.
Second, although Naive harms performance in prefetcher-adverse workloads, it significantly improves performance by $16.7\%$ in prefetcher-friendly workloads. 
Athena dynamically determines that enabling both POPET and Pythia is advantageous in these workloads, thus closely matching Naive's performance.
Overall, Athena outperforms Naive, HPAC, and MAB by $5.7\%$, $7.9\%$, and $5.0\%$, respectively, across \emph{all} $100$ workloads, demonstrating \emph{consistent} performance improvements.

\begin{figure}[!ht]
    \centering
    \includegraphics[width=\columnwidth]{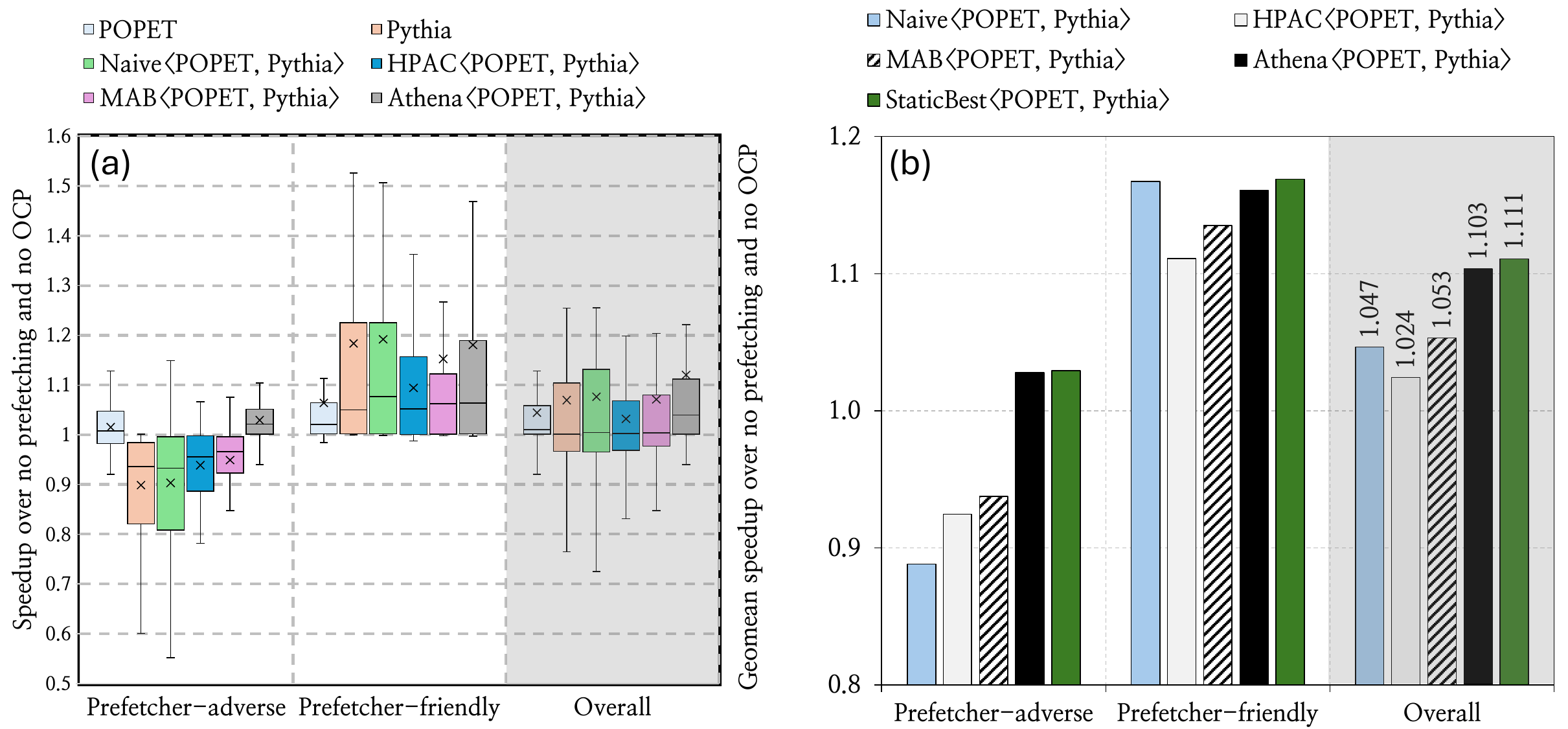}
    \caption{(a) Workload category-wise performance analysis in CD1. (b) Performance comparison with StaticBest in CD1.}
\label{fig:cd1_perf_deepdive}
\end{figure}

\paraheading{Workload Category-Wise Performance Analysis.}
To further analyze Athena's performance gains in CD1,~\Cref{fig:cd1_perf_deepdive}(a) shows the workload category-wise speedups as a box-and-whisker plot.
We make three key observations.
First, for prefetcher-adverse workloads, Athena substantially raises the lower quartile as well as the lower-end whisker relative to Naive, HPAC, and MAB.
This implies that Athena improves performance broadly across all prefetcher-adverse workloads, rather than merely alleviating a small number of extreme slowdowns.
Second, for prefetcher-friendly workloads, Athena increases both the upper quartile and the upper-end whisker compared to HPAC and MAB, demonstrating consistent performance improvements over these policies across all prefetcher-friendly workloads.
Third, when considering all workloads together, Athena improves mean performance relative to HPAC and MAB while simultaneously elevating both the lower and upper quartiles. 
This result indicates that Athena delivers robust and consistent performance gains across a broad spectrum of workload behaviors.

\paraheading{Performance Comparison with StaticBest.}
\Cref{fig:cd1_perf_deepdive}(b) compares the performance of Athena against the StaticBest combination (see~\Cref{subsubsec:naive_combo}). 
The key takeaway is that, by dynamically coordinating POPET and Pythia, Athena provides similar performance gains as the StaticBest combination for both prefetcher-adverse and prefetcher-friendly workload categories. 
On average, Athena improves performance by $10.3\%$ over the baseline with no prefetching or OCP, whereas StaticBest improves performance by $11.1\%$.

\subsubsection{\textbf{CD2: OCP with One L1D Prefetcher}}\label{subsubsec:L1D}
\begin{figure}[!ht]
    \centering
    \includegraphics[width=\columnwidth]{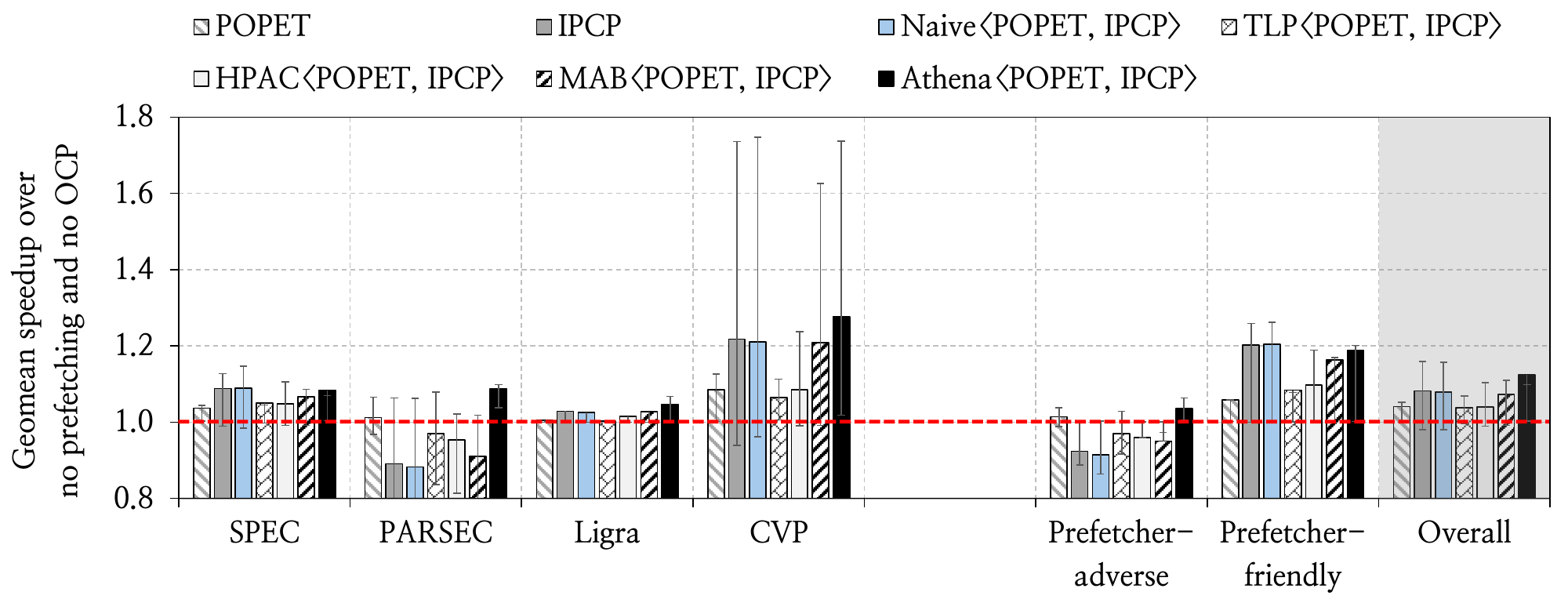}
    \caption{Speedup in cache design 2 (CD2).}
\label{fig:cd2}
\end{figure}

\Cref{fig:cd2} shows the performance improvement of Naive, TLP, HPAC, MAB, and Athena when coordinating POPET as the OCP and IPCP as the L1D prefetcher.
We make two observations. 
First, TLP outperforms Naive by $5.5\%$ in prefetcher-adverse workloads by filtering out prefetches that are predicted to go off-chip.
However, this filtering strategy hurts performance in prefetcher-friendly workloads where prefetch requests are indeed helpful, causing TLP to underperform Naive by $12.0\%$.
Second, Athena, by dynamically learning using multiple system-level features, outperforms TLP in both prefetcher-adverse and prefetcher-friendly workloads by $6.5\%$ and $10.4\%$, respectively, highlighting its robust and \emph{consistent} performance improvements.
Overall, Athena outperforms Naive, TLP, HPAC, and MAB on average by $4.5\%$, $8.7\%$, $8.4\%$, and $5.2\%$, respectively.

\subsubsection{\textbf{CD3: OCP with Two L2C Prefetchers}} \label{subsubsec:2L2C}
\begin{figure}[!ht]
    \centering
    \includegraphics[width=\columnwidth]{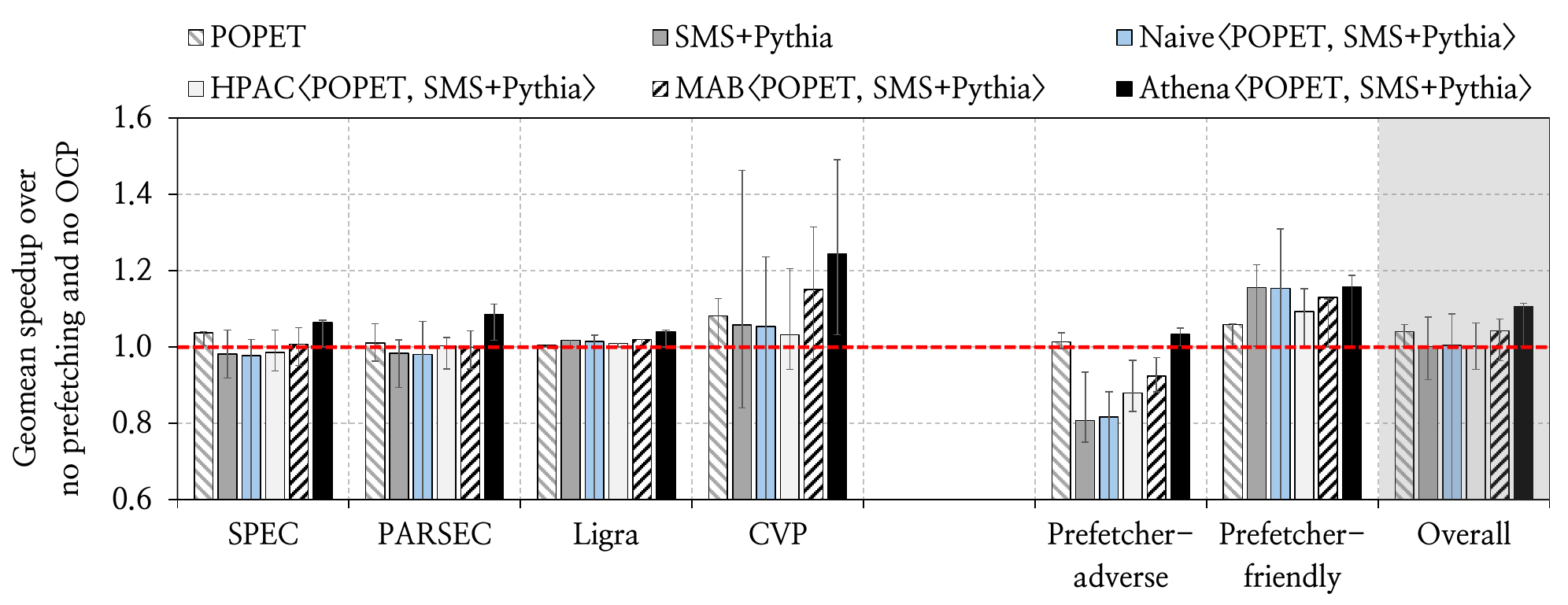}
    \caption{Speedup in cache design 3 (CD3).}
\label{fig:cd3}
\end{figure}

\Cref{fig:cd3} shows the performance improvement of Naive, HPAC, MAB, and Athena when coordinating POPET as the OCP, along with SMS and Pythia as two L2C prefetchers.
We observe that, in prefetcher-adverse workloads, HPAC and MAB only partially alleviate Naive's performance degradation, failing to match the baseline with no prefetching or OCP.
Athena, in contrast, achieves a $3.2\%$ improvement over the baseline, surpassing POPET's standalone performance.
In prefetcher-friendly workloads, Athena matches Naive's performance.
Overall, Athena outperforms Naive, HPAC, and MAB on average by $10.1\%$, $10.4\%$, and $6.4\%$, respectively, underscoring Athena's effectiveness irrespective of the cache design.

\subsubsection{\textbf{CD4: OCP with One L1D and One L2C Prefetcher}} \label{subsubsec:L1DL2C}
\begin{figure}[!ht]
    \centering
    \includegraphics[width=\columnwidth]{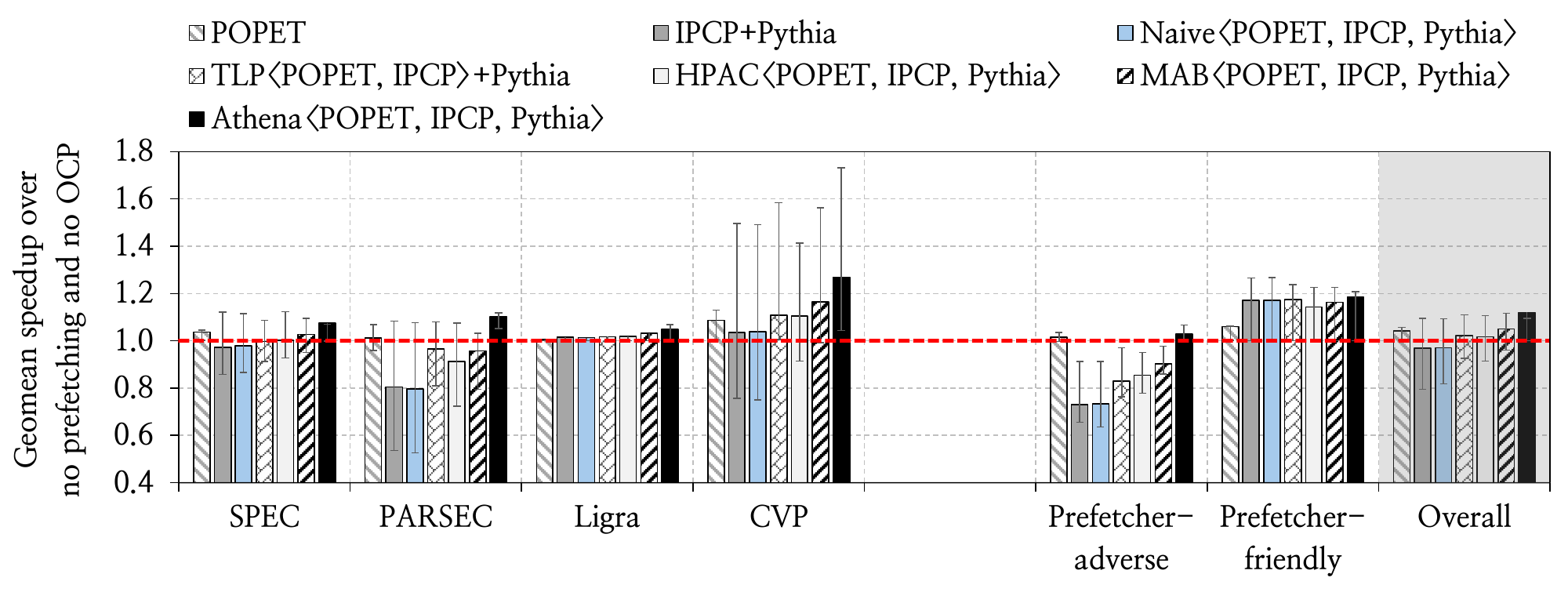}
    \caption{Speedup in cache design 4 (CD4).}
\label{fig:cd4}
\end{figure}

\Cref{fig:cd4} shows the performance improvement of Naive, TLP, HPAC, MAB, and Athena when coordinating POPET as the OCP, IPCP as the L1D prefetcher, and Pythia as the L2C prefetcher.
We make two key observations. 
First, in prefetcher-adverse workloads, enabling both prefetchers and the OCP without coordination results in a severe $26.8\%$ performance degradation, the worst among all evaluated cache designs.
TLP, due to its lack of control over the L2C prefetcher, fails to throttle harmful L2C prefetch requests, resulting in a performance degradation of $16.7\%$. 
Athena, benefiting from its flexibility, effectively coordinates prefetchers across two cache levels, significantly outperforming TLP by $19.9\%$.
Second, in prefetcher-friendly workloads, TLP closely matches Naive's performance since it inherently has no control over the L2C prefetcher.
In contrast, Athena dynamically determines that prefetching is beneficial, thereby providing higher performance compared to TLP.
Overall, Athena outperforms Naive, TLP, HPAC, and MAB on average by $14.9\%$, $9.9\%$, $10.3\%$, and $7.0\%$, respectively. 

Based on our extensive evaluation, we conclude that Athena \emph{consistently} outperforms prior coordination mechanisms (e.g., TLP, HPAC, and MAB) across diverse cache designs that employ OCP with multiple prefetchers at different cache levels.

\begin{figure*}[!ht]
    \centering
    \includegraphics[width=\textwidth]{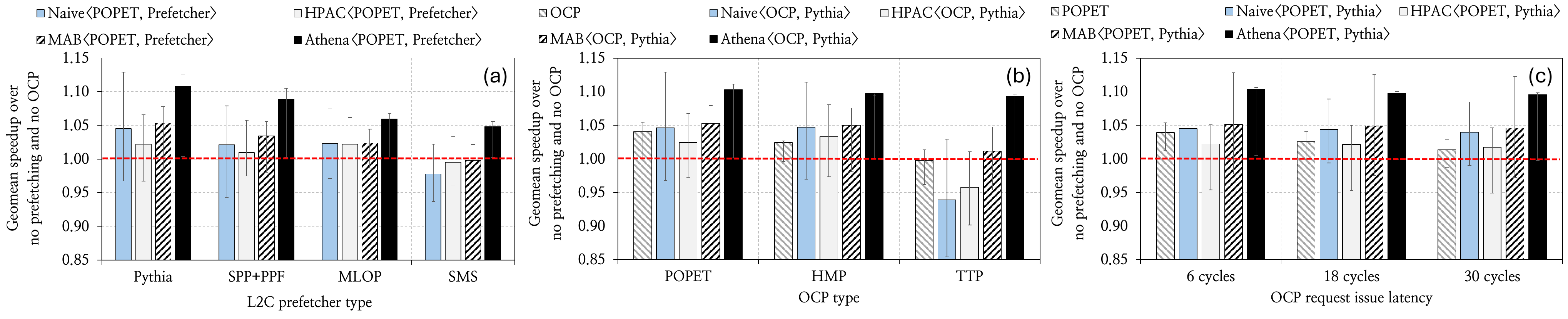}
    \caption{Performance sensitivity to (a) prefetching mechanisms at L2C, (b) off-chip prediction mechanisms, and (c) off-chip predicted request issue latency in CD1.}
\label{fig:cd1_sen_master}
\end{figure*}

\subsection{Performance Sensitivity Analysis in CD1} \label{sec:eval_sen_cd1}

While~\Cref{sec:eval_sc} demonstrates the benefits of Athena across various cache designs, this section further demonstrates Athena's adaptability by fixing the cache design to CD1 and varying the underlying L2C prefetcher and OCP type.

\subsubsection{\textbf{Effect of L2C Prefetcher Type}}
\Cref{fig:cd1_sen_master}(a) shows the performance improvement of Naive, HPAC, MAB, and Athena across all workloads while using POPET as the OCP, and varying the underlying L2C prefetcher.
The key observation is that, by autonomously learning using system-level features and telemetry information, Athena \emph{consistently} outperforms Naive, HPAC, and MAB for \emph{every} prefetcher type, without requiring any changes to its configuration. 
On average, Athena outperforms the next-best-performing MAB by $5.0$\%, $5.4$\%, $3.6$\%, and $5.0$\%, when coordinating POPET with four types of L2C prefetchers, Pythia, SPP+PPF, MLOP, and SMS, respectively.
We conclude that Athena is able to adapt and provide consistent performance benefits across diverse prefetcher types.

\subsubsection{\textbf{Effect of Off-Chip Predictor Type}} \label{sec:eval_sen_ocp}
\Cref{fig:cd1_sen_master}(b) shows the performance improvement of Naive, HPAC, MAB, and Athena while using Pythia as the L2C prefetcher and varying the underlying OCP.
The key takeaway is that Athena \emph{consistently} outperforms Naive, HPAC, and MAB for \emph{every} OCP type. 
On average, Athena outperforms the next-best-performing MAB by $5.0\%$, $4.7\%$, and $8.2\%$, when employing POPET, HMP, and TTP as the underlying OCP, respectively.
We conclude that Athena is able to adapt and provide consistent performance benefits across diverse OCP types.

\subsubsection{\textbf{Effect of OCP Request Issue Latency}} \label{subsubsec:ocp_latency}

For each load request predicted to go off-chip, an OCP issues a speculative memory request (we call it an \emph{OCP request}) directly to the main memory controller as soon as the physical address of the load becomes available. 
While an OCP request experiences substantially lower latency than a regular demand load, it still incurs a latency to traverse the on-chip network.
To faithfully evaluate OCP under a wide range of on-chip network designs, we vary the latency to directly issue an OCP request to the main memory controller (we call this \emph{OCP request issue latency}) from $6$ to $30$ cycles, in line with prior work~\cite{hermes}.

\Cref{fig:cd1_sen_master}(c) shows the performance improvement of Naive, HPAC, MAB, and Athena across all workloads when coordinating Pythia as the L2C prefetcher, POPET as the OCP, and varying the OCP request issue latency. 
We make three key observations.
First, POPET's performance gains decrease by $2.5\%$ as the OCP request issue latency increases from $6$ to $30$ cycles, consistent with prior work~\cite{hermes}.
Second, although the overall benefit of OCP reduces with higher request latency, Athena's performance decreases by only $0.8\%$, demonstrating robust adaptability to varying request delays.
Third, Athena \emph{consistently} outperforms Naive, HPAC, and MAB for all evaluated OCP request issue latencies.
We conclude that Athena is able to adapt to diverse system configurations with variations in on-chip network design.

\subsection{Performance Sensitivity Analysis in CD4} \label{sec:eval_sen_cd4}

This section further demonstrates Athena's adaptability in CD4, which employs one prefetcher each at L1D and L2C, by varying the L1D prefetcher type and main memory bandwidth.

\subsubsection{\textbf{Effect of L1D Prefetcher Type}} \label{subsubsec:eva_sen_l1d}
\begin{figure}[!ht]
    \centering
    \includegraphics[width=\columnwidth]{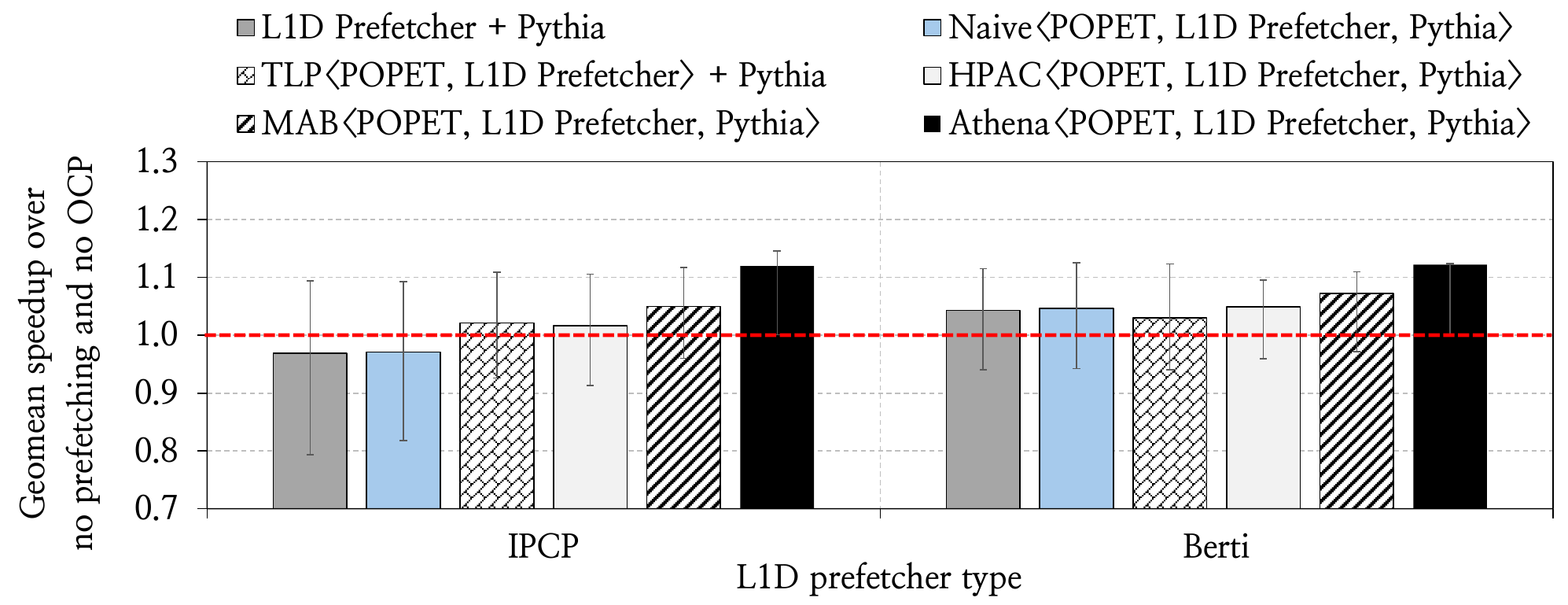}
    \caption{Performance sensitivity to prefetching mechanism at L1D in CD4.}
\label{fig:cd4_sen_1}
\end{figure}

\Cref{fig:cd4_sen_1} shows the performance improvement of Naive, TLP, HPAC, MAB, and Athena across all workloads while varying the underlying L1D prefetcher, but keeping POPET as the OCP and Pythia as the L2C prefetcher. 
We make two key observations.
First, Berti, due to its higher prefetch accuracy, provides a higher performance gain than IPCP.
Berti improves performance by $4.3\%$ on average over the baseline without any prefetcher or OCP, whereas IPCP degrades performance by $3.1\%$. 
Second, Athena \emph{consistently} improves performance, both over the baseline and prior coordination techniques, irrespective of the L1D prefetcher type.
Athena outperforms the next-best-performing MAB by $7.0\%$ and $5.0\%$ on average, while coordinating IPCP and Berti at L1D, respectively.

\subsubsection{\textbf{Effect of Main Memory Bandwidth}} \label{sec:eval_sen_bw}
\begin{figure}[!ht]
    \centering
    \includegraphics[width=\columnwidth]{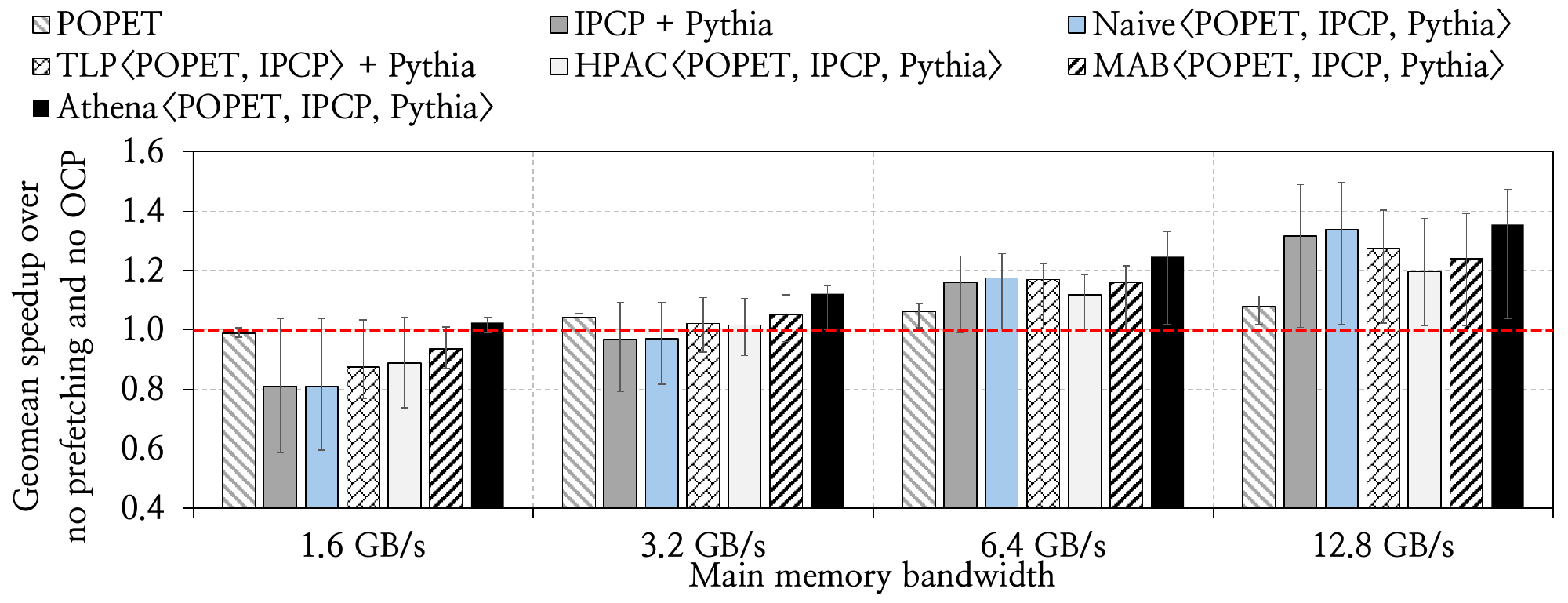}
    \caption{Performance sensitivity to main memory bandwidth in CD4.}
\label{fig:cd4_sen_2}
\end{figure}

\Cref{fig:cd4_sen_2} shows the performance improvement of Naive, TLP, HPAC, MAB, and Athena over the baseline with no prefetcher or OCP across all workloads, while varying the main memory bandwidth (measured in gigabytes per second (GB/s)).

We make three key observations.
First, while Naive significantly improves performance over the baseline in a system with ample main memory bandwidth, Naive's performance gain, which is largely dominated by the prefetchers, significantly deteriorates when the system has limited main memory bandwidth, akin to datacenter-class processors~\cite{epyc_9754, ampere_one, graviton3, bruce2023arm, neoverse2}.
For example, Naive \emph{improves} performance by $33.5\%$ on average in the system with $12.8$~GB/s main memory bandwidth, while it \emph{degrades} performance by $18.9\%$ in the system with $1.6$~GB/s main memory bandwidth.
Second, OCP alone also hurts performance in severely bandwidth-limited configurations, despite its highly accurate predictions. 
POPET degrades performance by $1.1\%$ on average over the baseline in the system with $1.6$~GB/s main memory bandwidth.
This indicates that neither prefetching nor off-chip prediction is beneficial for performance in severely bandwidth-constrained configurations.
Third, by autonomously learning using system-level features such as bandwidth usage, Athena \emph{consistently} outperforms Naive, TLP, HPAC, and MAB across \emph{all} bandwidth configurations. 
Athena's benefit is more prominent in bandwidth-constrained configurations since no static combination (i.e., POPET-alone, Pythia-alone, naive combination of POPET and Pythia, or none) is consistently good across all phases of all workloads.
However, in a system with ample bandwidth, Naive often yields good performance, and Athena correctly identifies this combination as the best-performing.
Overall, Athena outperforms Naive (MAB) by $21.4\%$ ($8.8\%$) and $1.7\%$ ($11.6\%$) in $1.6$~GB/s and $12.8$~GB/s bandwidth configurations, respectively.

\subsection{Multi-Core Evaluation Overview} \label{sec:eval_mc}
\subsubsection{\textbf{Four-Core Performance Analysis}}\label{subsubsec:4c}

\begin{figure}[!ht]
    \centering
    \includegraphics[width=\columnwidth]{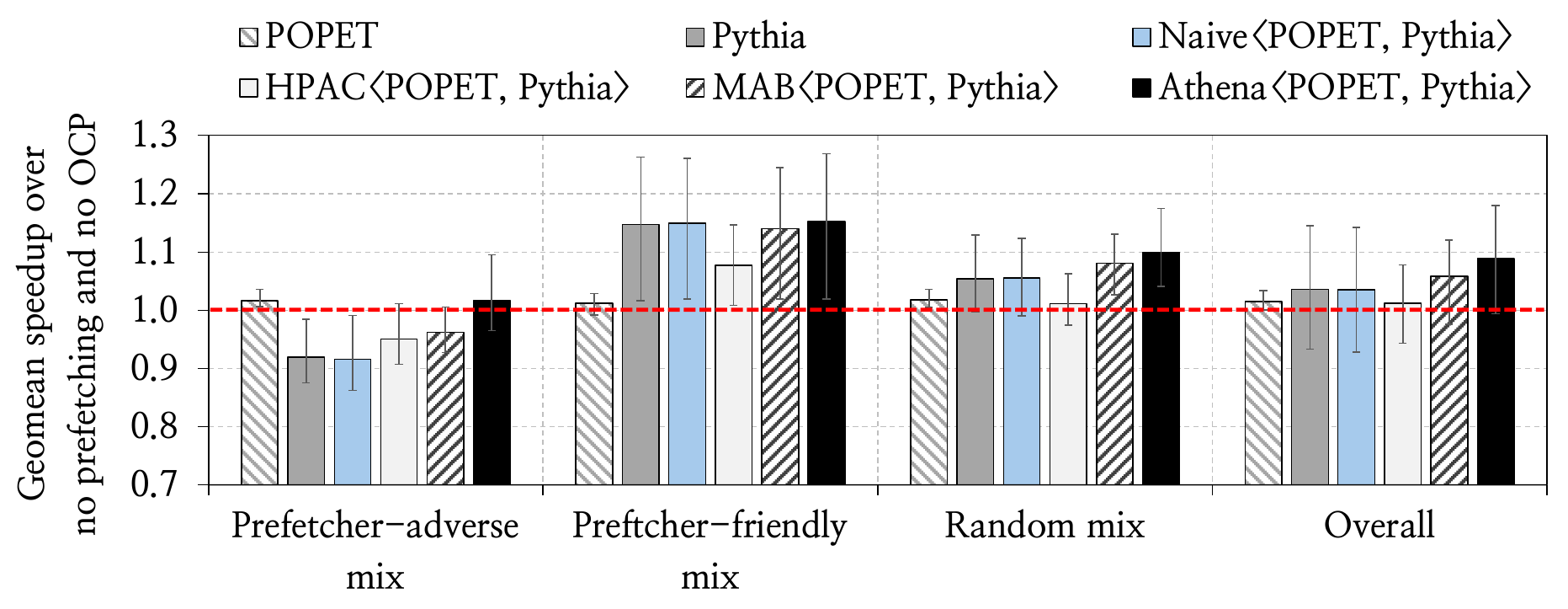}
    \caption{Speedup in four-core workloads.}
    \label{fig:perf_4c}
\end{figure}

\Cref{fig:perf_4c} shows the performance improvement of Naive, HPAC, MAB, and Athena when coordinating POPET as the OCP and Pythia as the L2C prefetcher in four-core workloads.

We make three key observations. 
First, across all workload mixes, Athena outperforms Naive, HPAC, and MAB by $5.3\%$, $7.7\%$, and $3.0\%$, respectively, despite utilizing hyperparameters exclusively tuned for single-core workloads (i.e., hyperparameters derived from the automated DSE described in \Cref{subsec:automated} are applied directly without alteration). 
Second, Athena \emph{consistently} outperforms \emph{all} prior coordination mechanisms in \emph{every} workload mix category.
Athena's performance gains over Naive, HPAC, and MAB are largest in prefetcher-adverse mixes, where no static configuration (i.e., POPET-alone, Pythia-alone, naive combination of POPET and Pythia, or none) is consistently good across all workload mixes. 
In contrast, Athena's relative benefit is smallest in prefetcher-friendly mixes, where the Naive combination is often beneficial for performance.
More specifically, Athena outperforms Naive (MAB) by $10.1\%$ ($5.5\%$), $0.4\%$ ($1.2\%$), and $4.5\%$ ($1.9\%$) on average in prefetcher-adverse, prefetcher-friendly, and random workload mixes, respectively. 
These results highlight Athena's ability to adaptively select effective coordination policies across diverse multi-core workload compositions.

\subsubsection{\textbf{Eight-Core Performance Analysis}}
\begin{figure}[!ht]
    \centering
    \includegraphics[width=\columnwidth]{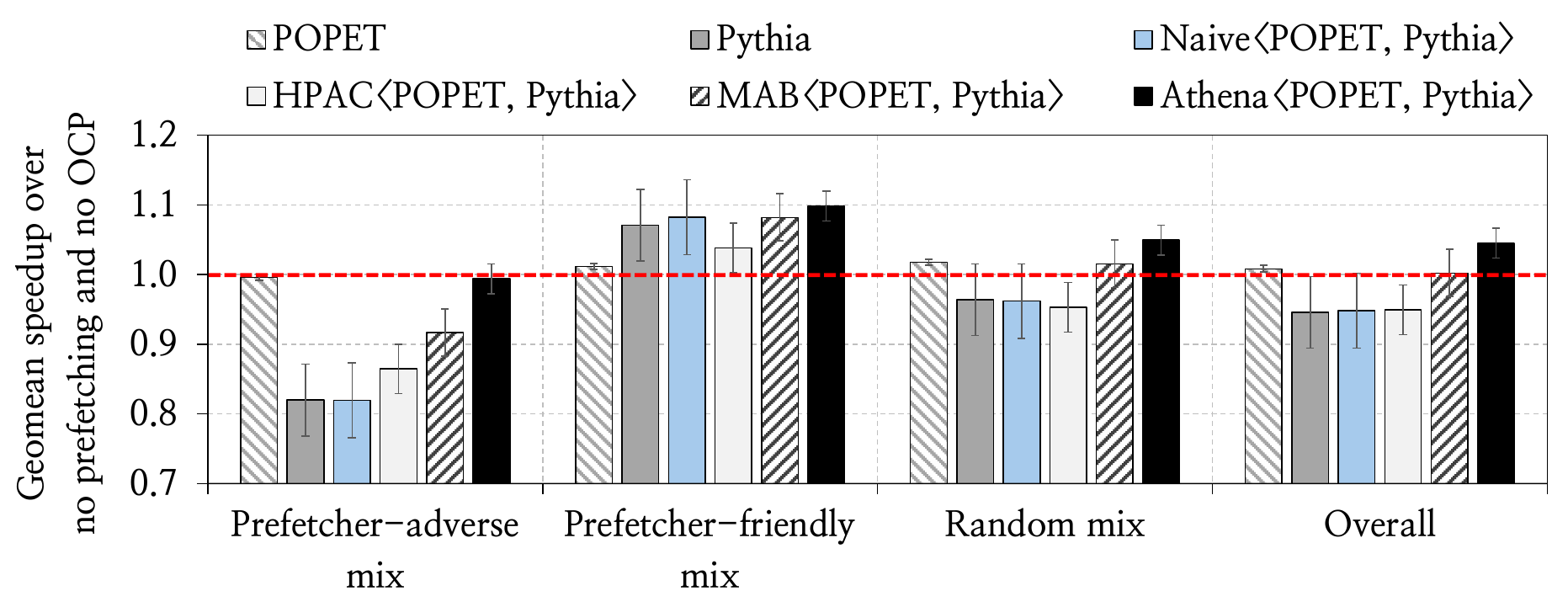}
    \caption{Speedup in eight-core workloads.}
    \label{fig:8c}
\end{figure}

\Cref{fig:8c} shows the performance improvement of Naive, HPAC, MAB, and Athena when coordinating POPET as the OCP and Pythia as the L2C prefetcher in eight-core workloads.
We highlight two key observations.
First, similar to four-core mixes, Athena consistently outperforms Naive, HPAC, and MAB by $9.7\%$, $9.6\%$, and $4.3\%$, respectively, across all eight-core mixes, despite using hyperparameters that are exclusively tuned for single-core workloads.
Second, Athena \emph{consistently} outperforms \emph{all} prior coordination mechanisms in \emph{every} workload mix category.

The multi-core results demonstrate that Athena surpasses existing coordination policies in multi-core settings, without requiring workload-specific tuning. We believe Athena could improve even further by tuning it specifically for multi-core.

\subsection{Understanding Athena} \label{subsec:deepdive_athenad}

\subsubsection{Understanding Athena's Decision-Making using a Case Study} \label{sec:eval_case_study}

\begin{figure}[!ht]
    \centering
    \includegraphics[width=\columnwidth]{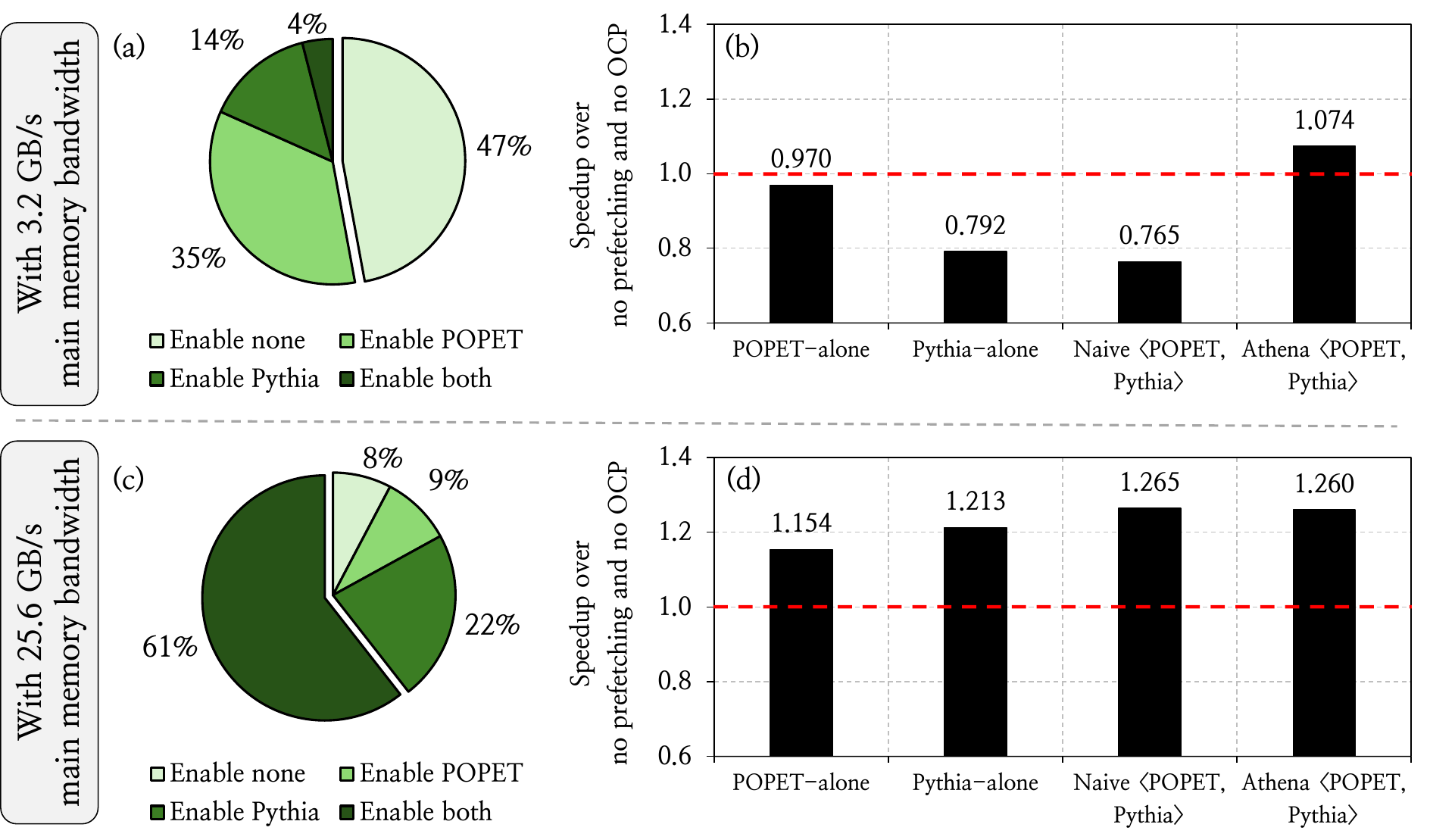}
    \caption{Distribution of Athena's action in coordinating Pythia and POPET and speedup of different Pythia-POPET combinations in \texttt{compute\_fp\_78} workload from \texttt{CVP} suite, while varying memory bandwidth: 3.2~GB/s and 25.6~GB/s.} 
    \label{fig:athena_deepdive1}
\end{figure}

To provide deeper insights into Athena's decision-making process, we analyze its actions in coordinating POPET as the OCP and Pythia as the L2C prefetcher in a representative workload, \texttt{compute\_fp\_78}, from the \texttt{CVP} suite. 
As~\Cref{fig:athena_deepdive1}(a) shows, in the system with $3.2$~GB/s main memory bandwidth, Athena disables both POPET and Pythia, or enables only POPET in $47\%$ and $35\%$ of its actions, respectively.
It enables only Pythia or both mechanisms in only $14\%$ and $4\%$ of its actions.
To find the rationale behind this action distribution, we independently evaluate the performance of three static combinations: POPET-alone, Pythia-alone, and Naive, for this workload. 
\Cref{fig:athena_deepdive1}(b) shows both Pythia-alone and Naive substantially degrade performance in the $3.2$~GB/s bandwidth configuration. 
While POPET-alone also incurs performance degradation, it does so less severely.
By selectively enabling POPET in specific epochs and disabling both mechanisms in most others, Athena effectively \emph{outperforms} all three static combinations.

When we evaluate the same workload in the system with $25.6$~GB/s main memory bandwidth, we observe that the action distribution significantly changes, and Athena strongly favors enabling \emph{both} Pythia and POPET.
As~\Cref{fig:athena_deepdive1}(c) shows, Athena enables both Pythia and POPET in $61\%$ of its actions.
The performance graph in~\Cref{fig:athena_deepdive1}(d) shows that, unlike in the $3.2$~GB/s configuration, both Pythia-alone and Naive significantly \emph{improve} performance.
We conclude that Athena is not only dynamically learning to find the best coordination, but also adapting to system configuration changes.

\subsubsection{Understanding the Sources of Athena's Performance Gains via Ablation Study} \label{subsec:eval_ablation}
\begin{figure}[!ht]
\centering
\includegraphics[width=\linewidth]{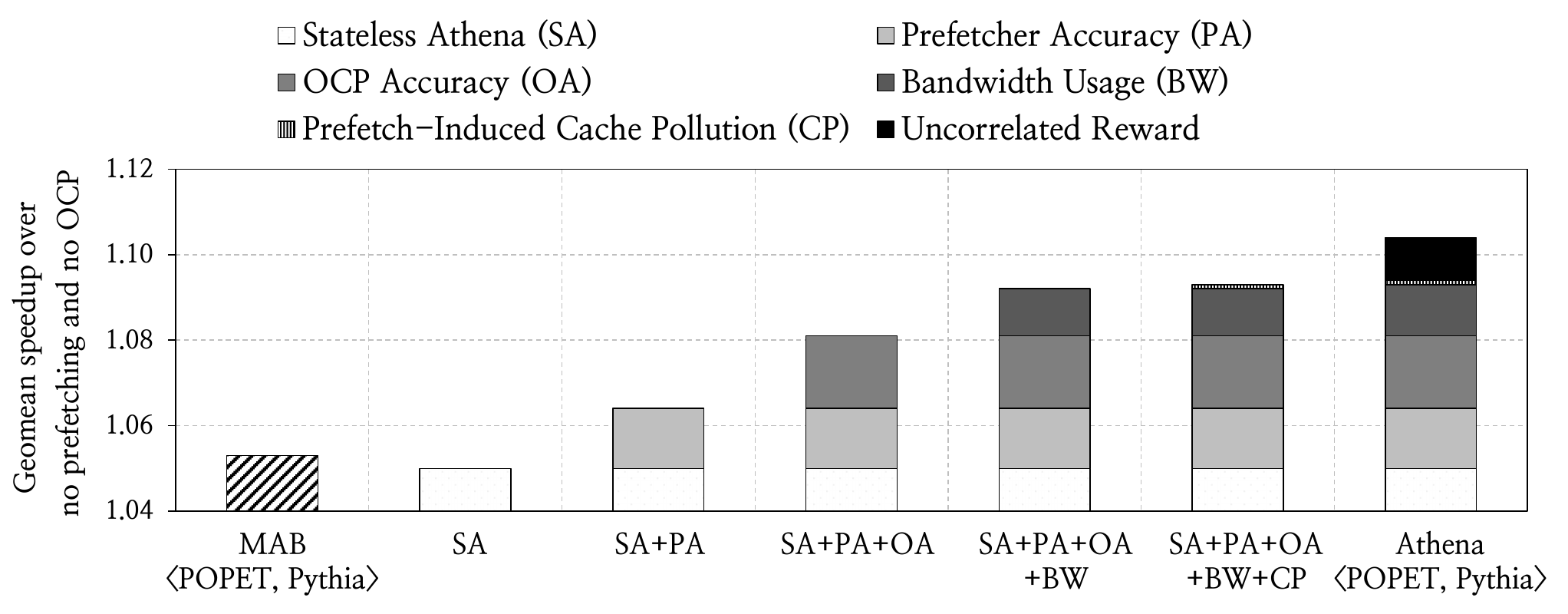}
\caption{Contribution of individual state features and reward components to Athena's geomean performance \edit[3]{across all 100 workloads}.}
\label{fig:ablation}
\end{figure}

To better understand the source of Athena's performance gains, we conduct an ablation study to evaluate the contribution of each state feature and reward component to overall performance. 
We begin with a version of Athena that uses no state information and employs only IPC as the correlated reward (see~\ref{subsec:reward}). 
We call this configuration \emph{Stateless Athena}.
We then progressively introduce each state feature (i.e., prefetcher accuracy, OCP accuracy, bandwidth utilization, and prefetch-induced cache pollution), and finally include the uncorrelated reward \edit[3]{(see~\ref{subsec:reward})}.
\Cref{fig:ablation} illustrates the effect of each state feature and reward component on Athena's geomean performance. 
Each bar represents Athena's performance up to and including the added state feature or reward component.

We make three key observations from~\Cref{fig:ablation}.
First, the stateless Athena, which operates without any state information, performs slightly worse than MAB, consistent with prior work~\cite{mab}. This result stems from the difference in exploration strategies: MAB employs Discounted Upper Confidence Bound (DUCB), whereas Athena uses $\epsilon$-greedy. In the stateless configuration, $\epsilon$-greedy selects random actions uniformly, with a non-decaying exploration rate, leading to slightly lower efficiency.
Second, incorporating prefetcher accuracy, OCP accuracy, bandwidth utilization, and prefetch-induced cache pollution progressively improves performance by $1.4\%$, $1.7\%$, $0.8\%$, and $0.1\%$, respectively, relative to the preceding configuration. However, as discussed in \Cref{subsec:automated}, adding additional features beyond prefetch-induced cache pollution yields diminishing returns. Hence, we limit Athena's state to four features.
Finally, adding the uncorrelated reward further improves performance by $1.0\%$, highlighting that the uncorrelated reward component significantly helps improve performance by isolating the true impact of Athena's actions from inherent variations in the workload.

\subsection{\textbf{Athena for Prefetcher-Only Management}} \label{subsubsec:athena_pref_only}

\begin{figure}[!ht]
    \centering
    \includegraphics[width=\columnwidth]{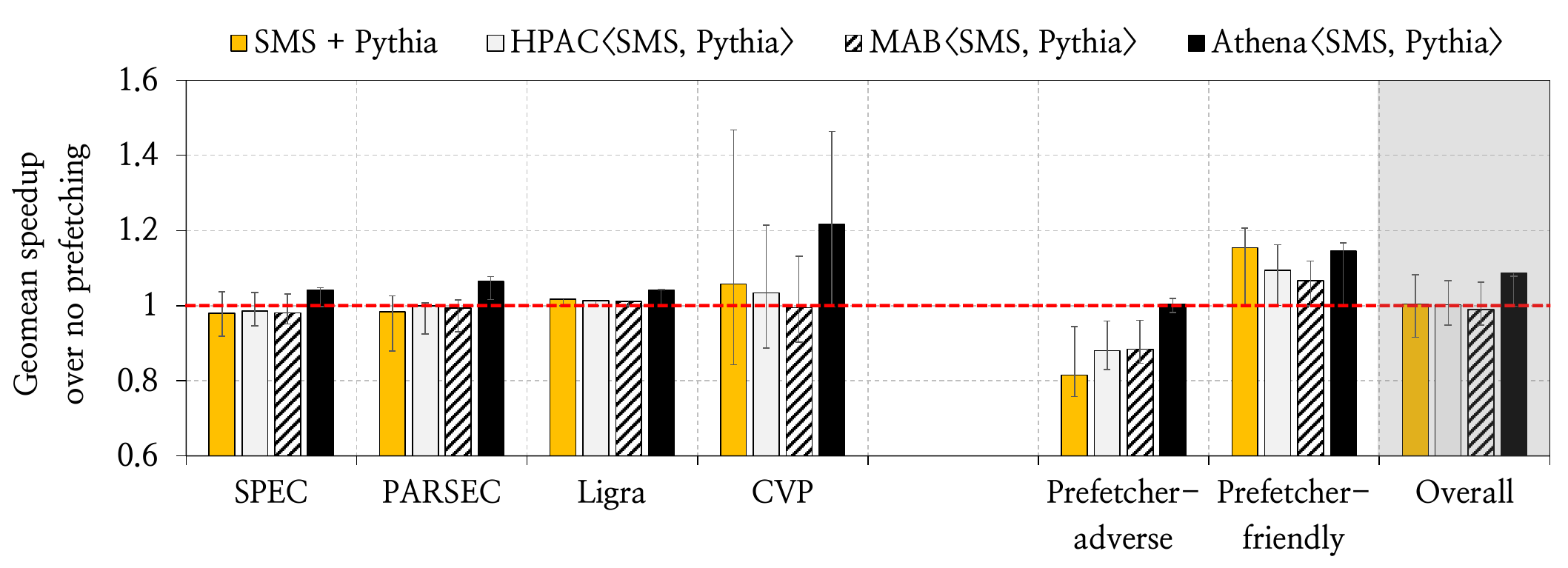}
    \caption{Geomean performance of Athena coordinating two L2C prefetchers without OCP.}
    \label{fig:generalizability}
\end{figure}

To evaluate the generality of Athena and its applicability beyond OCP-enabled systems, we conduct a generalizability study by comparing Athena against HPAC and MAB in a cache hierarchy without an OCP. 
Specifically, we evaluate Athena using a configuration that employs SMS and Pythia at the L2C (similar to CD3 in \Cref{subsubsec:2L2C}, but without OCP).
\Cref{fig:generalizability} shows the geomean performance of Naive, HPAC, MAB, and Athena, when coordinating SMS and Pythia as the L2C prefetchers.

We make two key observations.
First, in prefetcher-adverse workloads, both HPAC and MAB fail to adequately throttle prefetching, leading to performance degradation below the baseline with no prefetching. In contrast, Athena effectively mitigates these slowdowns, maintaining performance close to the baseline. However, in the absence of an OCP, which could otherwise act as a complementary mechanism, Athena only prevents performance loss as opposed to improving performance, as observed in \Cref{subsubsec:2L2C}.
Second, in prefetcher-friendly workloads, Athena consistently outperforms HPAC and MAB by $5.1\%$ and $7.8\%$, respectively. 
Overall, Athena achieves $7.6\%$ and $8.8\%$ higher performance than HPAC and MAB, respectively.
We conclude that Athena generalizes across system configurations with multiple prefetchers and maintains its adaptability even when the alternative mechanism, OCP, is absent.

\section{Related Works}

To our knowledge, Athena is the first work that proposes a reinforcement learning (RL)-based framework that coordinates OCP with multiple data prefetchers employed at various levels of the cache hierarchy.
We already quantitatively compared Athena against TLP~\cite{jamet2024tlp}, HPAC~\cite{ebrahimi2009coordinated}, and MAB~\cite{mab}.
In this section, we qualitatively compare Athena against other related works.

\paraheading{Data Prefetching.}
Prior prefetching techniques can be broadly categorized into three classes:
(1) precomputation-based prefetchers that pre-execute program code to generate prefetch addresses~\cite{dundas, mutlu2003runahead, mutlu2003runahead2, mutlu2005techniques, mutlu2006efficient, hashemi2016continuous, mutlu2005address, hashemi2015filtered, vector_runahead, mutlu2005reusing, iacobovici2004effective, decoupled_vector_runahead, roelandts2024svr, precise_runahead, naithani2022rar},
(2) temporal prefetchers that predict future load addresses by memorizing long sequences of demanded cacheline addresses~\cite{markov, stems, somogyi_stems, wenisch2010making, domino, isb, misb, triage, triangel, chilimbi2002dynamic, chou2007low, ferdman2007last, hu2003tcp, bekerman1999correlated, cooksey2002stateless, karlsson2000prefetching}, and 
(3) spatial prefetchers that predict future load addresses by learning program access patterns over different memory regions~\cite{stride, streamer, baer2, jouppi_prefetch, ampm, fdp, footprint, sms, spp, vldp, sandbox, bop, dol, dspatch, bingo, mlop, ppf, ipcp, kpc, pythia, chen2025gaze, dmp, navarro2022berti, navarro2025complexity, vavouliotis2025cross, vavouliotis2022ppm}. 
Athena can be combined with any prefetcher and provides \textit{consistent} performance benefits.

\paraheading{Machine Learning (ML) in Computer Architecture.}
Researchers have proposed ML-based algorithms for various microarchitectural decision-making. 
Prominent examples include memory scheduling~\cite{rlmc, morse}, 
cache management~\cite{glider, imitation, rl_cache, teran2016perceptron, jimenez2016multiperspective, balasubramanian2021accelerating, yang2023adaptive, chrome}, 
hybrid memory/storage management~\cite{singh2022sibyl, rakesh2025harmonia, chang2024idt}, 
branch prediction~\cite{tarsa, branchnet, perceptron, garza2019bit, BranchNetV2_2020, zouzias2021branchpa, jimenez2016multiperspective, tarjan2005merging, jimenez2003fast, jimenez2002neural}, 
address translation~\cite{margaritovTranslationNIPS18}, and 
hardware prefetching~\cite{peled2018neural, peled_rl, hashemi2018learning, shi2019learning, shineural, zeng2017long, long2023deep, pathfinder, zhang2022resemble, wu2023prefetching, mohapatra2023drishyam, huang2023rlop, duong2024new}. 
Researchers have also explored ML techniques to explore the large microarchitectural design space, e.g., NoC design~\cite{fettes2018dynamic, zheng2019energy, lin2020deep, yin2020experiences, ebrahimi2012haraq, ditomaso2016dynamic, clark2018lead, ditomaso2017machine, van2018extending, yin2018toward}, chip placement optimization~\cite{mirhoseini2021,mirhoseini2018hierarchical,mirhoseini2017device}, and hardware resource assignment and task allocation~\cite{kao2020confuciux, dubach2010predictive, ganapathi2009case, gomez2001neuro,jain2016machine, lu2015reinforcement, wu2012inferred, esteves2018adaptive, yazdanbakhsh2021apollo}. 
These works are orthogonal to Athena.

\section{Conclusion}

We introduce Athena, a reinforcement learning (RL)-based technique that synergizes data prefetchers and off-chip predictor (OCP) by autonomously learning from system behavior. 
Athena measures multiple system-level features (e.g., prefetcher/OCP accuracy, memory bandwidth usage) and uses them as state information to take an action: enabling the prefetcher and/or OCP, and adjusting the prefetcher aggressiveness. 
Athena introduces a holistic reward framework that disentangles events correlated to its own actions (e.g., improvement in IPC) from the events that are uncorrelated to its actions (e.g., change in mispredicted branch instructions). 
This allows Athena to autonomously learn a coordination policy by isolating the true impact of its actions from inherent variations in the workload.
Our extensive evaluation using a wide variety of workloads shows that Athena \emph{consistently} outperforms a naive prefetcher-OCP combination, heuristic-based HPAC~\cite{ebrahimi2009coordinated}, and learning-based TLP~\cite{jamet2024tlp} and MAB~\cite{mab}, across a wide range of system configurations with various combinations of underlying prefetchers at various cache levels, OCPs, and main memory bandwidths, while incurring only modest storage overhead.
We hope that Athena and its novel reward policy would inspire future works on data-driven coordination policy design. 
Such techniques would not only improve system performance and efficiency under a wide range of configurations, but would also reduce an architect's burden in designing sophisticated control policies.
\section*{Acknowledgments}
We thank the anonymous reviewers of MICRO 2025 and HPCA 2026 for their feedback.
We thank all SAFARI Research Group members for providing a stimulating \edit[3]{and inclusive} intellectual \edit[3]{and scientific} environment.
We acknowledge the generous gifts from our industrial partners: Futurewei, Google, Huawei, Intel, Microsoft, and VMware.
This work is supported in part by the Semiconductor Research Corporation and the ETH Future Computing Laboratory.
Rahul thanks his departed father, whom he misses dearly every day.

\bibliographystyle{IEEEtranS}
\bibliography{refs}

\appendix
\pagebreak
\section{Artifact Appendix}


\subsection{Abstract}

We evaluate Athena using the ChampSim trace-driven simulator. 
This artifact contains 
(1) the source code of Athena, 
(2) the source code of multiple data prefetching and off-chip prediction (OCP) mechanisms used to evaluate Athena, 
(3) the source code of multiple prior coordination mechanisms to compare against Athena, 
and (4) all necessary scripts to reproduce key performance results.\footnote{While this appendix focuses on reproducing nine key results, the artifact contains all files and scripts needed to reproduce all results presented in the paper.}

\vspace{2pt}
\noindent
We identify nine results that demonstrate Athena's novelty:
\begin{enumerate}
    \item \Cref{fig:cd1}: speedup in cache design 1 (CD1).
    \item \Cref{fig:cd2}: speedup in cache design 2 (CD2).
    \item \Cref{fig:cd3}: speedup in cache design 3 (CD3).
    \item \Cref{fig:cd4}: speedup in cache design 4 (CD4).
    \item \Cref{fig:cd1_sen_master}(a): Performance sensitivity to the L2C prefetcher in CD1.
    \item \Cref{fig:cd1_sen_master}(b): Performance sensitivity to the off-chip predictor (OCP) in CD1.
    \item \Cref{fig:cd1_sen_master}(c): Performance sensitivity to the OCP request issue latency in CD1.
    \item \Cref{fig:cd4_sen_1}: Performance sensitivity to the L1D prefetcher in CD4.
    \item \Cref{fig:cd4_sen_2}: Performance sensitivity to main memory bandwidth in CD4.
\end{enumerate}

\subsection{Artifact check-list (meta-information)}

{\small
\begin{itemize}
  \item {\bf Algorithm: } Reinforcement learning-based prefetcher coordination (Athena), heuristic-based coordination (HPAC), and learning-based coordination (TLP, MAB).
  \item {\bf Program: } ChampSim-based CPU simulator with prefetcher and off-chip prediction implementations.
  \item {\bf Compilation: } C++11 with g++ (GCC 11.3.0 or later), Makefile-based build.
  \item {\bf Data set: } 100 workload traces: \texttt{SPEC CPU 2006/2017} (49 traces), \texttt{PARSEC} (13 traces), \texttt{Ligra} (13 traces), \texttt{CVP} (25 traces).
  \item {\bf Run-time environment: } Linux (tested on Ubuntu 22.04+), Python 3.8+
  \item {\bf Hardware: } x86\_64 CPU, minimum 8GB RAM recommended.
  \item {\bf Execution: } Trace-driven simulation, SLURM job scheduler supported.
  \item {\bf Metrics: } IPC (Instructions Per Cycle), Geometric mean speedup.
  \item {\bf Output: } Per-trace statistics files (.out), aggregated CSV files, and PNG visualization plots.
  \item {\bf Experiments: } 9 studies. Each study runs on 100 traces.
  \item {\bf How much disk space required (approximately)?: } $\sim$1GB for source code and experiment outputs, $\sim$30GB for workload traces.
  \item {\bf How much time is needed to prepare workflow (approximately)?: } 15-30 minutes for setting up the infrastructure, $\sim$1 hour for downloading the traces.
  \item {\bf How much time is needed to complete experiments (approximately)?: } $\sim$8 hours per figure using 100+ parallel jobs on a compute cluster; $\sim$1-2 weeks without parallel jobs.
  \item {\bf Publicly available?: } Yes
  \item {\bf Code licenses (if publicly available)?: } MIT.
  \item {\bf Data licenses (if publicly available)?: } Traces available from public benchmark suites (\texttt{SPEC}, \texttt{PARSEC}, \texttt{Ligra}, and \texttt{CVP}).
  \item {\bf Workflow automation framework used?: } Python-based scripts and SLURM job scheduler.
  \item {\bf Archived (provide DOI)?: \url{https://doi.org/10.5281/zenodo.17854634} }
\end{itemize}
}


\subsection{Description}

\subsubsection{How to Access}

The artifact is available at \url{https://github.com/CMU-SAFARI/Athena.git}.
The workload traces are available at \url{https://doi.org/10.5281/zenodo.17850673}.

\subsubsection{Hardware Dependencies}

\begin{itemize}
  \item x86\_64 CPU
  \item Minimum 8GB RAM 
  \item For full reproduction: access to a compute cluster with a SLURM scheduler is highly recommended
\end{itemize}

\subsubsection{Software Dependencies}

\begin{itemize}
  \item Linux operating system (tested on Ubuntu 22.04)
  \item GCC/G++ 11.3.0 or later with C++11 support
  \item Python 3.8+ with packages: numpy, pandas, matplotlib
\end{itemize}

\subsubsection{Data sets}

The evaluation uses 100 workload traces from four benchmark suites:
\begin{itemize}
  \item 49 traces from \texttt{SPEC CPU 2006/2017}
  \item 13 traces from \texttt{PARSEC}
  \item 13 traces from \texttt{Ligra}
  \item 25 traces from \texttt{CVP}
\end{itemize}

\subsubsection{Models}

The artifact includes the following prefetchers and off-chip predictors (OCP):
\begin{itemize}
  \item L1D Prefetchers: IPCP, Berti
  \item L2C Prefetchers: Pythia, SMS, SPP+PPF, MLOP
  \item Off-Chip Predictors: POPET, HMP, TTP
  \item Coordination policies: Athena (RL-based), TLP (Two Level Perceptron), HPAC (Hierarchical Prefetcher Aggressiveness Control), and MAB (Micro-Armed Bandit)
\end{itemize}


\subsection{Installation}
Please follow the README. The following summarizes the key steps.

\begin{enumerate}
\item Clone the repository:
    \begin{lstlisting}[style=shell]
    git clone 
        https://github.com/CMU-SAFARI/Athena.git
    cd Athena
    \end{lstlisting}

\item Set up the environment:
    \begin{lstlisting}[style=shell]
    source setvars.sh
    \end{lstlisting}
    This sets the \texttt{ATHENA\_HOME} environment variable required by all scripts.

\item Build the simulator:
    \begin{lstlisting}[style=shell]
    make clean
    make -j$(nproc)
    \end{lstlisting}

\item Verify the compilation:
    \begin{lstlisting}[style=shell]
    ls bin/champsim
    # Should show: bin/champsim
    \end{lstlisting}
    The build produces a single binary \texttt{bin/champsim} that supports all prefetcher and off-chip predictor configurations via command-line arguments.

\item Download the workload traces:
    \begin{lstlisting}[style=shell]
    curl -L "https://zenodo.org/api/records/
    17850673/files-archive" -o download.zip
    \end{lstlisting}

\item Unzip the workload traces:
    \begin{lstlisting}[style=shell]
    mkdir traces
    unzip download.zip -d traces
    \end{lstlisting}

\item Verify trace integrity:
    \begin{lstlisting}[style=shell]
    mv checksum.txt $ATHENA_HOME/traces
    cd $ATHENA_HOME/traces
    sha256sum -c checksum.txt
    \end{lstlisting}
\end{enumerate}


\subsection{Experiment workflow}
Before launching the experiments, please make sure that \texttt{DEFAULT\_NCORES}, \texttt{DEFAULT\_PARTITION}, and \texttt{DEFAULT\_HOSTNAME} in \texttt{scripts/config.py} are correctly set for the SLURM cluster. 
The scripts provide a unified interface for all experiment operations:

\begin{enumerate}
\item \textbf{Launch experiments} to reproduce a specific figure (e.g., Figure 7):
    \begin{lstlisting}[style=shell]
    cd scripts
    python athena.py -L Fig7
    \end{lstlisting}

\item \textbf{Summarize results} after experiments complete:
    \begin{lstlisting}[style=shell]
    python athena.py -S Fig7
    \end{lstlisting}
    
    This should generate \texttt{\$ATHENA\_HOME/experiments/
    results/Fig7.csv}

\item \textbf{Relaunch failed experiments} (if any):
    \begin{lstlisting}[style=shell]
    python athena.py -R Fig7
    \end{lstlisting}

\item \textbf{Visualize results}:
    \begin{lstlisting}[style=shell]
    python athena.py -V Fig7
    \end{lstlisting}
    
    This should generate \texttt{\$ATHENA\_HOME/experiments/
    results/Fig7.png}

\item Repeat steps 1-4 for other figures.
\end{enumerate}


\subsection{Evaluation and expected results}

After running experiments for each cache design, the visualization script should generate the bar plots showing geometric mean speedup by workload category (SPEC, PARSEC, Ligra, CVP, Prefetcher-adverse, Prefetcher-friendly, Overall). 
We will focus on \emph{\textbf{the overall speedup}} values in each bar plot to verify the result reproduction. 
Results should match the paper figures within $\pm$0.1\% due to randomness in the simulation.

\vspace{2pt}
\noindent \textbf{Expected key results:}

\begin{itemize}
    \item \Cref{fig:cd1}: POPET, Pythia, Naive, HPAC, MAB, and Athena should show the overall speedup of $1.0406$, $1.0423$, $1.0465$, $1.0242$, $1.0531$, and $1.1034$, respectively.
    \item \Cref{fig:cd2}: POPET, Pythia, Naive, TLP, HPAC, MAB, and Athena should show the overall speedup of $1.0407$, $1.0820$, $1.0793$, $1.0370$, $1.0400$, $1.0726$, and $1.1245$, respectively.
    \item \Cref{fig:cd3}: POPET, SMS+Pythia, Naive, HPAC, MAB, and Athena should show the overall speedup of $1.0404$, $1.0013$, $1.0050$, $1.0019$, $1.0425$, and $1.1060$, respectively.
    \item \Cref{fig:cd4}: POPET, IPCP+Pythia, Naive, TLP, HPAC, MAB, and Athena should show the overall speedup of $1.0408$, $0.9690$, $0.9709$, $1.0212$, $1.0166$, $1.0501$, and $1.1198$, respectively.
    \item \Cref{fig:cd1_sen_master}(a): Athena should consistently outperform all prior coordination mechanisms for all L2C prefetchers. For Pythia, SPP+PPF, MLOP, and SMS, Athena's geomean speedup should be $1.1078$, $1.0889$, $1.0597$, $1.0483$, respectively.
    \item \Cref{fig:cd1_sen_master}(b): Athena should consistently outperform all prior coordination mechanisms for all OCP request latencies. For $6$-, $18$-, and $30$-cycle OCP request latency, Athena's geomean speedup should be $1.1038$, $1.0978$, and $1.0958$, respectively.
    \item \Cref{fig:cd1_sen_master}(c): Athena should consistently outperform all prior coordination mechanisms for all OCP types. For POPET, HMP, and TTP, Athena's geomean speedup should be $1.1034$, $1.0973$, and $1.0937$, respectively.
    \item \Cref{fig:cd4_sen_1}: Athena should consistently outperform all prior coordination mechanisms for all L1D prefetchers. For IPCP and Berti, Athena's geomean speedup should be $1.1198$ and $1.1223$, respectively. 
    \item \Cref{fig:cd4_sen_2}: Athena should consistently outperform all prior coordination mechanisms for all main memory bandwidth configurations. For $1.6$ GB/s, $3.2$ GB/s, $6.4$ GB/s, and $12.8$ GB/s configurations, Athena's geomean speedup should be $1.0250$, $1.1224$, $1.2469$, and $1.3554$, respectively.
\end{itemize}










\subsection{Methodology}

\begin{itemize}
  \item \url{https://www.acm.org/publications/policies/artifact-review-and-badging-current}
  \item \url{https://cTuning.org/ae}
\end{itemize}

\ifarxiv
    \clearpage

\section{Extended Results}

\subsection{Effect of Athena on Main Memory Requests}

\Cref{fig:cd1_dram_llc}(a) shows the number of main memory requests issued by POPET-alone, Pythia-alone, Naive, HPAC, MAB, and Athena across all $100$ workloads in CD1.
We make two key observations.
First, in prefetcher-adverse workloads, 
Pythia-alone significantly increases the main memory requests due to its poor prefetch accuracy.
Naively combining POPET with Pythia further increases the main memory requests (by $46.5\%$ on top of the baseline system without any prefetcher or OCP).
This overhead in main memory requests severely harms the overall performance (see~\cref{subsubsec:L2C}).
Athena, by dynamically coordinating POPET and Pythia, substantially reduces the main memory request overhead (only $5.9\%$ over the baseline).
Second, across all workloads, Naive, HPAC, and MAB increase the main memory requests by $21.9\%$, $15.2\%$, and $12.7\%$, respectively, whereas Athena increases them by only $5.8\%$ over the baseline without any prefetcher or OCP.

\begin{figure}[!ht]
    \centering
    \includegraphics[width=\columnwidth]{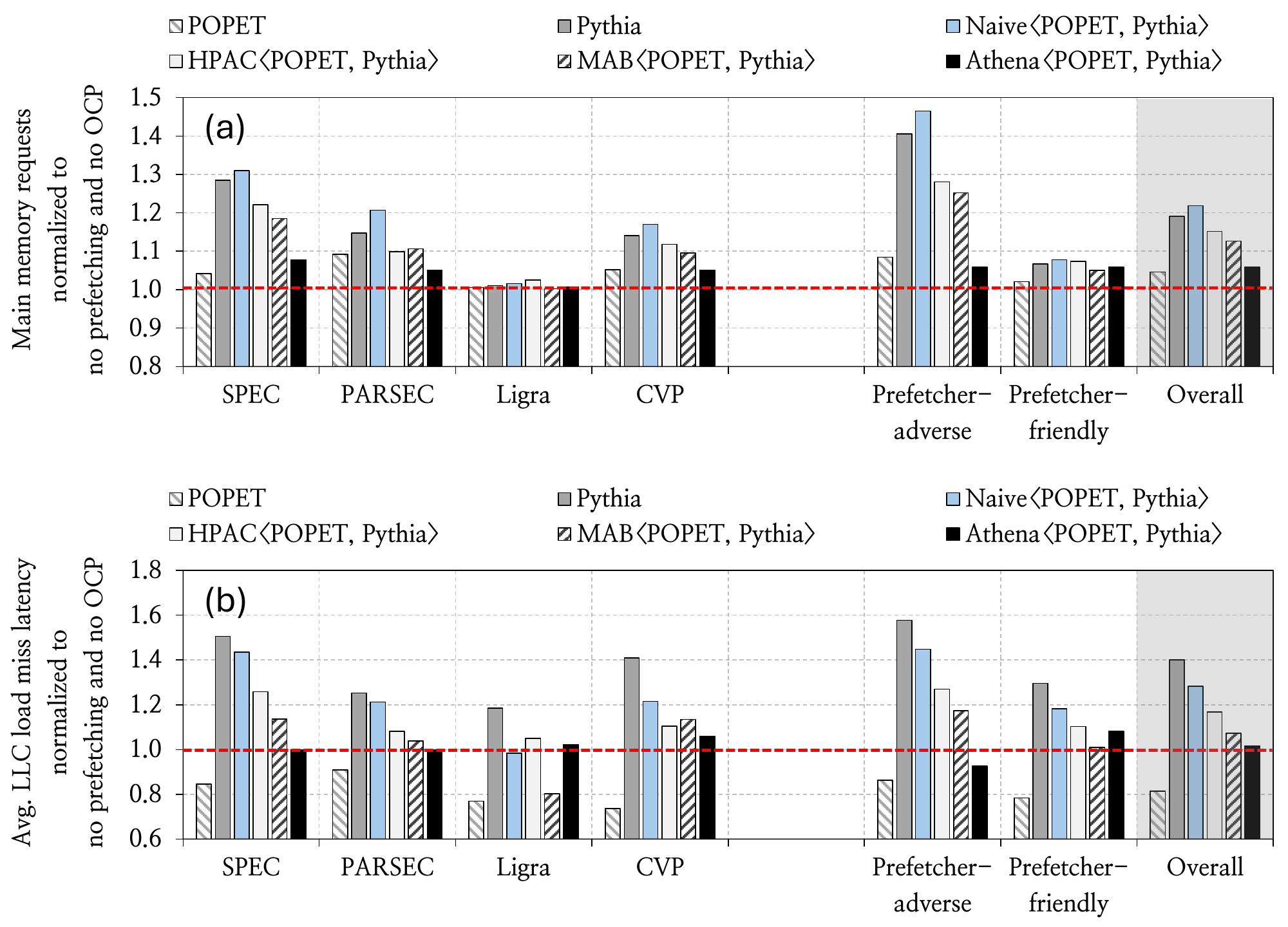}
    \caption{Comparison of (a) the number of main memory requests and (b) average last-level cache (LLC) load miss latency.}
\label{fig:cd1_dram_llc}
\end{figure}

\subsection{Effect of Athena on LLC Load Miss Latency}

\Cref{fig:cd1_dram_llc}(b) shows the average last-level cache (LLC) load miss latency, normalized to the baseline system without prefetching or OCP, across all $100$ workloads. 
We make three key observations. 
First, naively enabling both Pythia and POPET increases average LLC miss latency, particularly for prefetcher-adverse workloads, indicating that uncoordinated speculation can exacerbate memory-system contention and interference. 
Second, HPAC and MAB partially reduce this latency overhead, but still fall short of the baseline.
Third, Athena consistently reduces the LLC miss latency overhead across all workload categories by dynamically coordinating Pythia and POPET.
Overall, Naive, HPAC, and MAB increase the average LLC load miss latency by $28.3\%$, $16.7\%$, and $7.3\%$, respectively, whereas Athena increases it by only $1.7\%$ over the baseline without any prefetcher or OCP.

\subsection{Performance of Athena on Unseen Workloads}

To demonstrate Athena’s ability to provide performance gains across workload categories that are not used at all to tune Athena, we evaluate Athena using an additional set of $359$ traces from Google workloads released in the 4th Data Prefetching Championship (DPC4)~\cite{dpc4}.
These workload traces were originally captured from Google's warehouse-scale computers~\cite{gtrace_dynamorio} using the DynamoRIO instrumentation tool~\cite{dynamorio} and later converted for ChampSim. 
The traces are categorized into $12$ different groups.
We simulate these workloads using the same methodology described in~\Cref{sec:meth_workloads}.
\Cref{fig:cd4_unseen} shows the performance improvement of Naive, TLP, HPAC, MAB, and Athena, when coordinating POPET as the OCP, IPCP as the L1D prefetcher, and Pythia as the L2C prefetcher in CD4.
The key takeaway is that Athena outperforms the next-best performing coordination mechanism, TLP, across all workloads. 
Athena improves performance by $2.8\%$ on average over the baseline without any prefetcher or OCP, whereas MAB and TLP improve performance by $0.1\%$ and $2.2\%$, and HPAC and Naive degrade performance by $1.3\%$ and $2.1\%$, respectively.

\begin{figure}[!ht]
    \centering
    \includegraphics[width=\columnwidth]{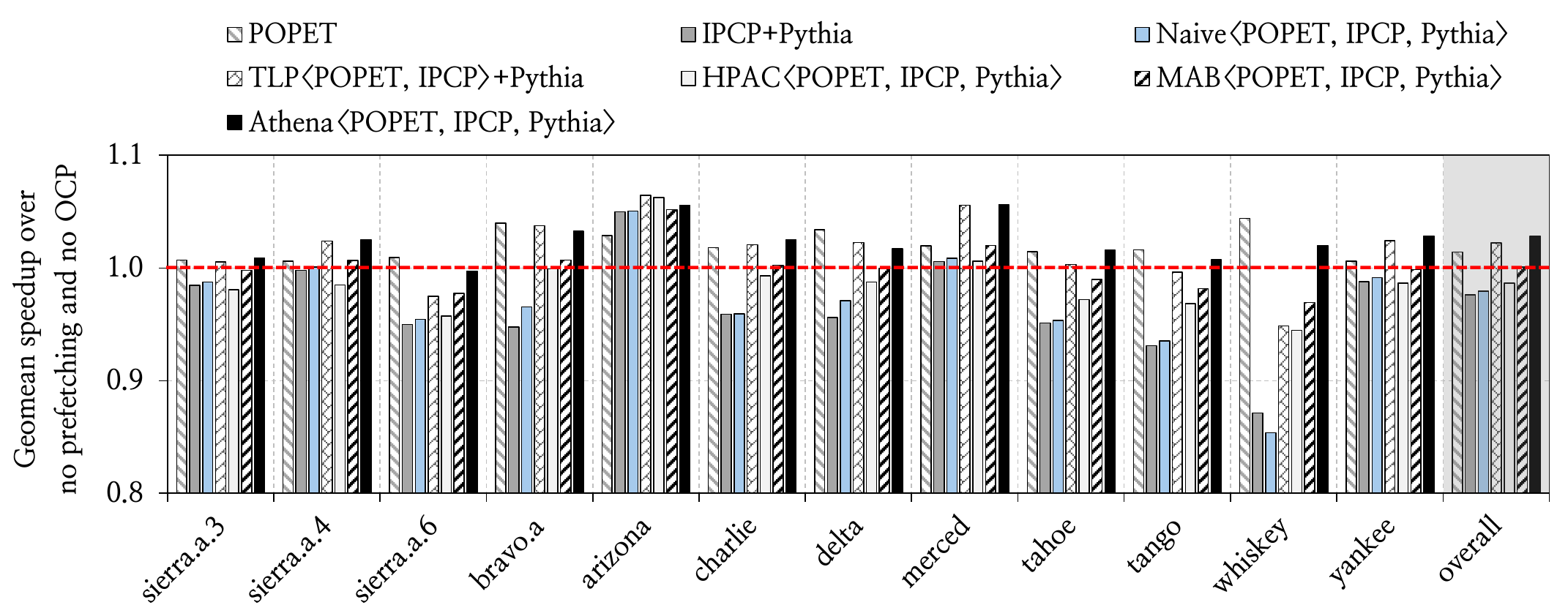}
    \caption{Speedup in unseen workload traces.}
\label{fig:cd4_unseen}
\end{figure}

\fi

\end{document}